%
%
%
\magnification = \magstep 1

\baselineskip=15pt plus 1pt minus 1pt
\font\secfnt=cmss10

\font\titlefnt=cmssbx10 scaled \magstep 1
\def\title#1{\centerline{{\titlefnt #1}}\medskip}

\def\author#1{\smallskip\centerline{#1}\medskip}
\def\address#1{\centerline{#1}}
\def\date#1{\smallskip\centerline{{\it #1}}\smallskip}

\def\abstract#1{\par\vskip\normalbaselineskip\par
    {\baselineskip=\normalbaselineskip
    \parindent=0 pt
    {\hfill\vbox{\hsize= 11 cm  #1  }\hfill}}
    \bigskip}

\def\section#1{\bigskip\centerline{{\secfnt #1}}\medskip}


\newcount\eqncnt
\eqncnt=0
\def\eqprefix{}
\def\eqn{\global\advance\eqncnt by 1 {\rm(\eqprefix\the\eqncnt)}}
\def\eqname#1{\eqn\xdef#1{\eqprefix\the\eqncnt}}


\newcount\refcnt
\refcnt=0
\def\ref#1.#2\par{\global\advance\refcnt by 1\xdef#1{\the\refcnt}}


\newcount\figcnt
\figcnt=0
\def\fig#1.#2\par{\global\advance\figcnt by 1\xdef#1{\the\figcnt}}

\def\ie{{\it i.e.}}

\def\etal{{\it et al}}



\def\threehalves{ {3\over 2} }
\def\fivehalves{ {5\over 2} }

\def\grapprox{\mathop{\lower.5ex \hbox{$\buildrel{\fivesy >}\over{\fivesy\sim}$}} \nolimits}
\def\lsapprox{\mathop{\lower.5ex \hbox{$\buildrel{\fivesy <}\over{\fivesy\sim}$}} \nolimits}
\def\grls{\mathop{\lower.5ex \hbox{$\buildrel{\fivesy >}\over{\fivesy <}$}} \nolimits}

\def\vec#1{{\bf #1}}

\def\avg#1{\left\langle #1 \right\rangle}
\def\abs#1{\left\vert #1 \right\vert}

\def\LBR{\left\lbrace}
\def\RBR{\right\rbrace}
\def\LB{\left\lbrack}
\def\RB{\right\rbrack}
\def\LP{\left (}
\def\RP{\right )}
\def\qq{\qquad\qquad}

\def\ptt#1{{\partial #1\over\partial t}}

\def\ppt#1{\partial #1/\partial t}

\def\grad{\nabla}
\def\cross{{\bf \times}}
\def\div{\grad\cdot}

\def\curl{\grad\cross}
\def\dpl{\grad_\parallel}
\def\ddpl{\grad_\parallel^2}
\def\dpp{\grad_\perp}
\def\ddpp{\grad_\perp^2}
\def\delsq{\grad^2}

\def\bunit{\vec{b}}

\def\fromto{\leftrightarrow}

\def\dotdot{\!:\!}

\def\bdel{\vec b\cdot\grad}

\def\jxb{\vec J\cross\vec B}

\def\vxb{\vec v\cross\vec B}

\def\ucxb{{\vec u\over c}\cross\vec B}
\def\vcxb{{\vec v\over c}\cross\vec B}

\def\jcxb{{\vec J\cross\vec B\over c}}

\def\vexb{\vec v_E}

\def\upol{\vec u_p}
\def\vstar{\vec v_*}
\def\ustar{\vec u_*}
\def\Jstar{\vec J_*}
\def\Jpol{\vec J_p}

\def\qestar{\vec q_e{}_\wedge}
\def\qistar{\vec q_i{}_\wedge}
\def\pistar{\vec\Pi_*}

\def\vdl{\vec v\cdot\grad}

\def\vedl{\vexb\cdot\grad}

\def\udel{\vec u\cdot\grad}

\def\jpp{J_\perp}
\def\jperp{\vec\jpp}

\def\upp{u_\perp}
\def\uperp{\vec\upp}

\def\vpp{v_\perp}
\def\vperp{\vec\vpp}

\def\Jpl{J_\parallel}

\def\Jpp{J_\perp}
\def\jpp{J_\perp}
\def\Jperp{\vec\Jpp}
\def\Bperp{\vec B_\perp}
\def\Apl{A_\parallel}

\def\App{A_\perp}

\def\Epl{E_\parallel}

\def\Epp{E_\perp}

\def\Eperp{\vec\Epp}
\def\upl{u_\parallel}
\def\vpl{v_\parallel}

\def\kpl{k_\parallel}

\def\kpp{k_\perp}

\def\Dpl{D_\parallel}
\def\Dpp{\Delta_\perp}

\def\qepl{q_e{}_\parallel}
\def\qipl{q_i{}_\parallel}
\def\mupl{\mu_\parallel}

\def\rs{\rho_s}

\def\rch{\rho_{ch}}
\def\npl{\eta_\parallel}

\def\npp{\eta_\perp}

\def\ptb{\widetilde}

\def\phifl{\widetilde\phi}

\def\nefl{\widetilde n_e}
\def\nifl{\widetilde n_i}
\def\tefl{\widetilde T_e}

\def\pfl{\widetilde p}
\def\pefl{\widetilde p_e}
\def\pifl{\widetilde p_i}

\def\vefl{\widetilde \vexb}

\def\Bfl{\widetilde \vec B}

\def\Appfl{\widetilde A_\perp}

\def\ufl{\widetilde u_\parallel}

\def\qefl{\widetilde q_e{}_\parallel}
\def\qifl{\widetilde q_i{}_\parallel}

\def\Afl{\ptb A_\parallel}
\def\Jfl{\ptb J_\parallel}

\def\qepl{ q_e{}_\parallel}

\def\qipl{ q_i{}_\parallel}


\def\prl#1{{\it Phys. Rev. Lett.} {\secfnt #1}}

\def\pf#1{{\it Phys. Fluids} {\secfnt #1}}

\def\pfb#1{{\it Phys. Fluids B} {\secfnt #1}}
\def\physp#1{{\it Phys. Plasmas} {\secfnt #1}}
\def\nf#1{{\it Nucl. Fusion} {\secfnt #1}}

\def\ppcf#1{{\it Plasma Phys. Contr. Fusion} {\secfnt #1}}

\def\revpp#1{{\it Rev. Plasma Phys.} {\secfnt #1}}

\def\aa#1{{\it Astron. Astrophys.} {\secfnt #1}}
\def\vol#1{\ {\secfnt #1}}

\def\temp{1.34}%
\let\tempp=\relax
\expandafter\ifx\csname psboxversion\endcsname\relax
  \message{PSBOX(\temp) loading}%
\else
    \ifdim\temp cm>\psboxversion cm
      \message{PSBOX(\temp) loading}%
    \else
      \message{PSBOX(\psboxversion) is already loaded: I won't load
        PSBOX(\temp)!}%
      \let\temp=\psboxversion
      \let\tempp= 
    \fi
\fi
\tempp
\let\psboxversion=\temp
\catcode`\@=11
%
%
\def\psfortextures{
\def\PSspeci@l##1##2{%
\special{illustration ##1\space scaled ##2}%
}}%
\def\psfordvitops{
\def\PSspeci@l##1##2{%
\special{dvitops: import ##1\space \the\drawingwd \the\drawinght}%
}}%
\def\psfordvips{
\def\PSspeci@l##1##2{%
\d@my=0.1bp \d@mx=\drawingwd \divide\d@mx by\d@my
\includegraphics{##1\space}}}%
\def\psforoztex{
\def\PSspeci@l##1##2{%
\special{##1 \space
      ##2 1000 div dup scale
      \number-\psllx\space \number-\pslly\space translate
}}}%
\def\psfordvitps{
\def\psdimt@n@sp##1{\d@mx=##1\relax\edef\psn@sp{\number\d@mx}}
\def\PSspeci@l##1##2{%
\special{dvitps: Include0 "psfig.psr"}
\psdimt@n@sp{\drawingwd}
\special{dvitps: Literal "\psn@sp\space"}
\psdimt@n@sp{\drawinght}
\special{dvitps: Literal "\psn@sp\space"}
\psdimt@n@sp{\psllx bp}
\special{dvitps: Literal "\psn@sp\space"}
\psdimt@n@sp{\pslly bp}
\special{dvitps: Literal "\psn@sp\space"}
\psdimt@n@sp{\psurx bp}
\special{dvitps: Literal "\psn@sp\space"}
\psdimt@n@sp{\psury bp}
\special{dvitps: Literal "\psn@sp\space startTexFig\space"}
\special{dvitps: Include1 "##1"}
\special{dvitps: Literal "endTexFig\space"}
}}%
\def\psfordvialw{
\def\PSspeci@l##1##2{
\special{language "PostScript",
position = "bottom left",
literal "  \psllx\space \pslly\space translate
  ##2 1000 div dup scale
  -\psllx\space -\pslly\space translate",
include "##1"}
}}%
\def\psforptips{
\def\PSspeci@l##1##2{{
\d@mx=\psurx bp
\advance \d@mx by -\psllx bp
\divide \d@mx by 1000\multiply\d@mx by \xscale
\incm{\d@mx}
\let\tmpx\dimincm
\d@my=\psury bp
\advance \d@my by -\pslly bp
\divide \d@my by 1000\multiply\d@my by \xscale
\incm{\d@my}
\let\tmpy\dimincm
\d@mx=-\psllx bp
\divide \d@mx by 1000\multiply\d@mx by \xscale
\d@my=-\pslly bp
\divide \d@my by 1000\multiply\d@my by \xscale
\at(\d@mx;\d@my){\special{ps:##1 x=\tmpx, y=\tmpy}}
}}}%
\def\psonlyboxes{
\def\PSspeci@l##1##2{%
\at(0cm;0cm){\boxit{\vbox to\drawinght
  {\vss\hbox to\drawingwd{\at(0cm;0cm){\hbox{({\tt##1})}}\hss}}}}
}}%
\def\psloc@lerr#1{%
\let\savedPSspeci@l=\PSspeci@l%
\def\PSspeci@l##1##2{%
\at(0cm;0cm){\boxit{\vbox to\drawinght
  {\vss\hbox to\drawingwd{\at(0cm;0cm){\hbox{({\tt##1}) #1}}\hss}}}}
\let\PSspeci@l=\savedPSspeci@l
}}%
%
%
\newread\pst@mpin
\newdimen\drawinght\newdimen\drawingwd
\newdimen\psxoffset\newdimen\psyoffset
\newbox\drawingBox
\newcount\xscale \newcount\yscale \newdimen\pscm\pscm=1cm
\newdimen\d@mx \newdimen\d@my
\newdimen\pswdincr \newdimen\pshtincr
\let\ps@nnotation=\relax
{\catcode`\|=0 |catcode`|\=12 |catcode`|
|catcode`#=12 |catcode`*=14
|xdef|backslashother{\}*
|xdef|percentother{
|xdef|tildeother{~}*
|xdef|sharpother{#}*
}%
\def\R@moveMeaningHeader#1:->{}%
\def\uncatcode#1{%
\edef#1{\expandafter\R@moveMeaningHeader\meaning#1}}%
\def\execute#1{#1}
\def\psm@keother#1{\catcode`#112\relax}
\def\executeinspecs#1{%
\execute{\begingroup\let\do\psm@keother\dospecials\catcode`\^^M=9#1\endgroup}}%
\def\@mpty{}%
\def\matchexpin#1#2{
  \fi%
  \edef\tmpb{{#2}}%
  \expandafter\makem@tchtmp\tmpb%
  \edef\tmpa{#1}\edef\tmpb{#2}%
  \expandafter\expandafter\expandafter\m@tchtmp\expandafter\tmpa\tmpb\endm@tch%
  \if\match%
}%
\def\matchin#1#2{%
  \fi%
  \makem@tchtmp{#2}%
  \m@tchtmp#1#2\endm@tch%
  \if\match%
}%
\def\makem@tchtmp#1{\def\m@tchtmp##1#1##2\endm@tch{%
  \def\tmpa{##1}\def\tmpb{##2}\let\m@tchtmp=\relax%
  \ifx\tmpb\@mpty\def\match{YN}%
  \else\def\match{YY}\fi%
}}%
\def\incm#1{{\psxoffset=1cm\d@my=#1
 \d@mx=\d@my
  \divide\d@mx by \psxoffset
  \xdef\dimincm{\number\d@mx.}
  \advance\d@my by -\number\d@mx cm
  \multiply\d@my by 100
 \d@mx=\d@my
  \divide\d@mx by \psxoffset
  \edef\dimincm{\dimincm\number\d@mx}
  \advance\d@my by -\number\d@mx cm
  \multiply\d@my by 100
 \d@mx=\d@my
  \divide\d@mx by \psxoffset
  \xdef\dimincm{\dimincm\number\d@mx}
}}%
%
\newif\ifNotB@undingBox
\newhelp\PShelp{Proceed: you'll have a 5cm square blank box instead of
your graphics (Jean Orloff).}%
\def\s@tsize#1 #2 #3 #4\@ndsize{
  \def\psllx{#1}\def\pslly{#2}%
  \def\psurx{#3}\def\psury{#4}
  \ifx\psurx\@mpty\NotB@undingBoxtrue
  \else
    \drawinght=#4bp\advance\drawinght by-#2bp
    \drawingwd=#3bp\advance\drawingwd by-#1bp
  \fi
  }%
\def\sc@nBBline#1:#2\@ndBBline{\edef\p@rameter{#1}\edef\v@lue{#2}}%
\def\g@bblefirstblank#1#2:{\ifx#1 \else#1\fi#2}%
{\catcode`\%=12
\xdef\B@undingBox{
\def\ReadPSize#1{
 \readfilename#1\relax
 \let\PSfilename=\lastreadfilename
 \openin\pst@mpin=#1\relax
 \ifeof\pst@mpin \errhelp=\PShelp
   \errmessage{I haven't found your postscript file (\PSfilename)}%
   \psloc@lerr{was not found}%
   \s@tsize 0 0 142 142\@ndsize
   \closein\pst@mpin
 \else
   \if\matchexpin{\GlobalInputList}{, \lastreadfilename}%
   \else\xdef\GlobalInputList{\GlobalInputList, \lastreadfilename}%
     \immediate\write\psbj@inaux{\lastreadfilename,}%
   \fi%
   \loop
     \executeinspecs{\catcode`\ =10\global\read\pst@mpin to\n@xtline}%
     \ifeof\pst@mpin
       \errhelp=\PShelp
       \errmessage{(\PSfilename) is not an Encapsulated PostScript File:
           I could not find any \B@undingBox: line.}%
       \edef\v@lue{0 0 142 142:}%
       \psloc@lerr{is not an EPSFile}%
       \NotB@undingBoxfalse
     \else
       \expandafter\sc@nBBline\n@xtline:\@ndBBline
       \ifx\p@rameter\B@undingBox\NotB@undingBoxfalse
         \edef\t@mp{%
           \expandafter\g@bblefirstblank\v@lue\space\space\space}%
         \expandafter\s@tsize\t@mp\@ndsize
       \else\NotB@undingBoxtrue
       \fi
     \fi
   \ifNotB@undingBox\repeat
   \closein\pst@mpin
 \fi
\message{#1}%
}%
%
%
\def\psboxto(#1;#2)#3{\vbox{
   \ReadPSize{#3}%
   \divide\drawingwd by 1000
   \divide\drawinght by 1000
   \d@mx=#1
   \ifdim\d@mx=0pt\xscale=1000
         \else \xscale=\d@mx \divide \xscale by \drawingwd\fi
   \d@my=#2
   \ifdim\d@my=0pt\yscale=1000
         \else \yscale=\d@my \divide \yscale by \drawinght\fi
   \ifnum\yscale=1000
         \else\ifnum\xscale=1000\xscale=\yscale
                    \else\ifnum\yscale<\xscale\xscale=\yscale\fi
              \fi
   \fi
   \divide\pswdincr by 1000 \multiply\pswdincr by \xscale
   \divide\pshtincr by 1000 \multiply\pshtincr by \xscale
   \divide\psxoffset by1000 \multiply\psxoffset by\xscale
   \divide\psyoffset by1000 \multiply\psyoffset by\xscale
   \global\divide\pscm by 1000
   \global\multiply\pscm by\xscale
   \multiply\drawingwd by\xscale \multiply\drawinght by\xscale
   \ifdim\d@mx=0pt\d@mx=\drawingwd\fi
   \ifdim\d@my=0pt\d@my=\drawinght\fi
   \message{scaled \the\xscale}%
 \hbox to\d@mx{\hss\vbox to\d@my{\vss
   \global\setbox\drawingBox=\hbox to 0pt{\kern\psxoffset\vbox to 0pt{
      \kern-\psyoffset
      \PSspeci@l{\PSfilename}{\the\xscale}%
      \vss}\hss\ps@nnotation}%
   \advance\pswdincr by \drawingwd
   \advance\pshtincr by \drawinght
   \global\wd\drawingBox=\the\pswdincr
   \global\ht\drawingBox=\the\pshtincr
   \baselineskip=0pt
   \copy\drawingBox
 \vss}\hss}%
  \global\psxoffset=0pt
  \global\psyoffset=0pt
  \global\pswdincr=0pt
  \global\pshtincr=0pt 
  \global\pscm=1cm 
  \global\drawingwd=\drawingwd
  \global\drawinght=\drawinght
}}%
%
%
\def\psboxscaled#1#2{\vbox{
  \ReadPSize{#2}%
  \xscale=#1
  \message{scaled \the\xscale}%
  \advance\drawingwd by\pswdincr\advance\drawinght by\pshtincr
  \divide\pswdincr by 1000 \multiply\pswdincr by \xscale
  \divide\pshtincr by 1000 \multiply\pshtincr by \xscale
  \divide\psxoffset by1000 \multiply\psxoffset by\xscale
  \divide\psyoffset by1000 \multiply\psyoffset by\xscale
  \divide\drawingwd by1000 \multiply\drawingwd by\xscale
  \divide\drawinght by1000 \multiply\drawinght by\xscale
  \global\divide\pscm by 1000
  \global\multiply\pscm by\xscale
  \global\setbox\drawingBox=\hbox to 0pt{\kern\psxoffset\vbox to 0pt{
     \kern-\psyoffset
     \PSspeci@l{\PSfilename}{\the\xscale}%
     \vss}\hss\ps@nnotation}%
  \advance\pswdincr by \drawingwd
  \advance\pshtincr by \drawinght
  \global\wd\drawingBox=\the\pswdincr
  \global\ht\drawingBox=\the\pshtincr
  \baselineskip=0pt
  \copy\drawingBox
  \global\psxoffset=0pt
  \global\psyoffset=0pt
  \global\pswdincr=0pt
  \global\pshtincr=0pt 
  \global\pscm=1cm
  \global\drawingwd=\drawingwd
  \global\drawinght=\drawinght
}}%
%
\def\psbox#1{\psboxscaled{1000}{#1}}%
\newif\ifn@teof\n@teoftrue
\newif\ifc@ntrolline
\newif\ifmatch
\newread\j@insplitin
\newwrite\j@insplitout
\newwrite\psbj@inaux
\immediate\openout\psbj@inaux=psbjoin.aux
\immediate\write\psbj@inaux{\string\joinfiles}%
\immediate\write\psbj@inaux{\jobname,}%
%
%
\def\toother#1{\ifcat\relax#1\else\expandafter%
  \toother@ux\meaning#1\endtoother@ux\fi}%
\def\toother@ux#1 #2#3\endtoother@ux{\def\tmp{#3}%
  \ifx\tmp\@mpty\def\tmp{#2}\let\next=\relax%
  \else\def\next{\toother@ux#2#3\endtoother@ux}\fi%
\next}%
%
%
\let\readfilenamehook=\relax
\def\re@d{\expandafter\re@daux}
\def\re@daux{\futurelet\nextchar\stopre@dtest}%
\def\re@dnext{\xdef\lastreadfilename{\lastreadfilename\nextchar}%
  \afterassignment\re@d\let\nextchar}%
\def\stopre@d{\egroup\readfilenamehook}%
\def\stopre@dtest{%
  \ifcat\nextchar\relax\let\nextread\stopre@d
  \else
    \ifcat\nextchar\space\def\nextread{%
      \afterassignment\stopre@d\chardef\nextchar=`}%
    \else\let\nextread=\re@dnext
      \toother\nextchar
      \edef\nextchar{\tmp}%
    \fi
  \fi\nextread}%
\def\readfilename{\vbox\bgroup%
  \let\\=\backslashother \let\%=\percentother \let\~=\tildeother
  \let\#=\sharpother \xdef\lastreadfilename{}%
  \re@d}%
%
%
\xdef\GlobalInputList{\jobname}%
\def\psnewinput{%
  \def\readfilenamehook{
    \if\matchexpin{\GlobalInputList}{, \lastreadfilename}%
    \else\xdef\GlobalInputList{\GlobalInputList, \lastreadfilename}%
      \immediate\write\psbj@inaux{\lastreadfilename,}%
    \fi%
    \ps@ldinput\lastreadfilename\relax%
    \let\readfilenamehook=\relax%
  }\readfilename%
}%
\expandafter\ifx\csname @@input\endcsname\relax    
  \immediate\let\ps@ldinput=\input\def\input{\psnewinput}%
\else
  \immediate\let\ps@ldinput=\@@input
  \def\@@input{\psnewinput}%
\fi%
\def\nowarnopenout{%
 \def\warnopenout##1##2{%
   \readfilename##2\relax
   \message{\lastreadfilename}%
   \immediate\openout##1=\lastreadfilename\relax}}%
\def\warnopenout#1#2{%
 \readfilename#2\relax
 \def\t@mp{TrashMe,psbjoin.aux,psbjoint.tex,}\uncatcode\t@mp
 \if\matchexpin{\t@mp}{\lastreadfilename,}%
 \else
   \immediate\openin\pst@mpin=\lastreadfilename\relax
   \ifeof\pst@mpin
     \else
     \errhelp{If the content of this file is so precious to you, abort (ie
press x or e) and rename it before retrying.}%
     \errmessage{I'm just about to replace your file named \lastreadfilename}%
   \fi
   \immediate\closein\pst@mpin
 \fi
 \message{\lastreadfilename}%
 \immediate\openout#1=\lastreadfilename\relax}%
{\catcode`\%=12\catcode`\*=14
\gdef\splitfile#1{*
 \readfilename#1\relax
 \immediate\openin\j@insplitin=\lastreadfilename\relax
 \ifeof\j@insplitin
   \message{! I couldn't find and split \lastreadfilename!}*
 \else
   \immediate\openout\j@insplitout=TrashMe
   \message{< Splitting \lastreadfilename\space into}*
   \loop
     \ifeof\j@insplitin
       \immediate\closein\j@insplitin\n@teoffalse
     \else
       \n@teoftrue
       \executeinspecs{\global\read\j@insplitin to\spl@tinline\expandafter
         \ch@ckbeginnewfile\spl@tinline
       \ifc@ntrolline
       \else
         \toks0=\expandafter{\spl@tinline}*
         \immediate\write\j@insplitout{\the\toks0}*
       \fi
     \fi
   \ifn@teof\repeat
   \immediate\closeout\j@insplitout
 \fi\message{>}*
}*
\gdef\ch@ckbeginnewfile#1
 \def\t@mp{#1}*
 \ifx\@mpty\t@mp
   \def\t@mp{#3}*
   \ifx\@mpty\t@mp
     \global\c@ntrollinefalse
   \else
     \immediate\closeout\j@insplitout
     \warnopenout\j@insplitout{#2}*
     \global\c@ntrollinetrue
   \fi
 \else
   \global\c@ntrollinefalse
 \fi}*
\gdef\joinfiles#1\into#2{*
 \message{< Joining following files into}*
 \warnopenout\j@insplitout{#2}*
 \message{:}*
 {*
 \edef\w@##1{\immediate\write\j@insplitout{##1}}*
\w@{
\w@{
\w@{
\w@{
\w@{
\w@{
\w@{
\w@{
\w@{
\w@{
\w@{\string\input\space psbox.tex}*
\w@{\string\splitfile{\string\jobname}}*
\w@{\string\let\string\autojoin=\string\relax}*
}*
 \expandafter\tre@tfilelist#1, \endtre@t
 \immediate\closeout\j@insplitout
 \message{>}*
}*
\gdef\tre@tfilelist#1, #2\endtre@t{*
 \readfilename#1\relax
 \ifx\@mpty\lastreadfilename
 \else
   \immediate\openin\j@insplitin=\lastreadfilename\relax
   \ifeof\j@insplitin
     \errmessage{I couldn't find file \lastreadfilename}*
   \else
     \message{\lastreadfilename}*
     \immediate\write\j@insplitout{
     \executeinspecs{\global\read\j@insplitin to\oldj@ininline}*
     \loop
       \ifeof\j@insplitin\immediate\closein\j@insplitin\n@teoffalse
       \else\n@teoftrue
         \executeinspecs{\global\read\j@insplitin to\j@ininline}*
         \toks0=\expandafter{\oldj@ininline}*
         \let\oldj@ininline=\j@ininline
         \immediate\write\j@insplitout{\the\toks0}*
       \fi
     \ifn@teof
     \repeat
   \immediate\closein\j@insplitin
   \fi
   \tre@tfilelist#2, \endtre@t
 \fi}*
}%
\def\autojoin{%
 \immediate\write\psbj@inaux{\string\into{psbjoint.tex}}%
 \immediate\closeout\psbj@inaux
 \expandafter\joinfiles\GlobalInputList\into{psbjoint.tex}%
}%
%
%
%
\def\centinsert#1{\midinsert\line{\hss#1\hss}\endinsert}%
\def\psannotate#1#2{\vbox{%
  \def\ps@nnotation{#2\global\let\ps@nnotation=\relax}#1}}%
\def\pscaption#1#2{\vbox{%
   \setbox\drawingBox=#1
   \copy\drawingBox
   \vskip\baselineskip
   \vbox{\hsize=\wd\drawingBox\setbox0=\hbox{#2}%
     \ifdim\wd0>\hsize
       \noindent\unhbox0\tolerance=5000
    \else\centerline{\box0}%
    \fi
}}}%
%
\def\at(#1;#2)#3{\setbox0=\hbox{#3}\ht0=0pt\dp0=0pt
  \rlap{\kern#1\vbox to0pt{\kern-#2\box0\vss}}}%
%
\newdimen\gridht \newdimen\gridwd
\def\gridfill(#1;#2){%
  \setbox0=\hbox to 1\pscm
  {\vrule height1\pscm width.4pt\leaders\hrule\hfill}%
  \gridht=#1
  \divide\gridht by \ht0
  \multiply\gridht by \ht0
  \gridwd=#2
  \divide\gridwd by \wd0
  \multiply\gridwd by \wd0
  \advance \gridwd by \wd0
  \vbox to \gridht{\leaders\hbox to\gridwd{\leaders\box0\hfill}\vfill}}%
%
\def\fillinggrid{\at(0cm;0cm){\vbox{%
  \gridfill(\drawinght;\drawingwd)}}}%
%
%
\def\textleftof#1:{%
  \setbox1=#1
  \setbox0=\vbox\bgroup
    \advance\hsize by -\wd1 \advance\hsize by -2em}%
\def\textrightof#1:{%
  \setbox0=#1
  \setbox1=\vbox\bgroup
    \advance\hsize by -\wd0 \advance\hsize by -2em}%
\def\endtext{%
  \egroup
  \hbox to \hsize{\valign{\vfil##\vfil\cr%
\box0\cr%
\noalign{\hss}\box1\cr}}}%
%
\def\frameit#1#2#3{\hbox{\vrule width#1\vbox{%
  \hrule height#1\vskip#2\hbox{\hskip#2\vbox{#3}\hskip#2}%
        \vskip#2\hrule height#1}\vrule width#1}}%
\def\boxit#1{\frameit{0.4pt}{0pt}{#1}}%
\catcode`\@=12 
%
 \psfordvips   

\def\figturbtrans{1}
\def\figetrans{2}

\baselineskip 15 pt plus 2 pt minus 1 pt

\parskip 6 pt

\def\ref#1{[#1]}
\def\cite#1{[#1]}

\def\Dpl{\Delta_\parallel}
\def\Dpp{\Delta_\perp}

\def\Bppfl{\Bfl_\perp}
\def\Appfl{\ptb \App}

\def\drift{{c\over B^2}\,\vec B\cross}
\def\mdrift{{M_i c\over Z e B^2}\,\vec B\cross}

\def\Jpol{\vec J_p}

\def\Jfl{\ptb\Jpl}

\def\vv{\vec v}
\def\uu{\vec u}
\def\xx{\vec x}
\def\ww{\vec w}

\def\ee{\vec E}
\def\bb{\vec B}
\def\jj{\vec J}
\def\qq{\vec q}
\def\aa{\vec A}

\def\GG{\vec g}

\def\Bfl{\ptb B}
\def\bbfl{\ptb\bb}
\def\jjfl{\ptb\jj}

\def\bunitfl{\ptb\bunit}
\def\upfl{\ptb\upol}
\def\vefl{\ptb\vexb}
\def\Bperp{\bb_\perp}
\def\bbfl{\ptb\bunit_\perp}
\def\uppfl{\ptb\upp}

\def\tei{\vec T_{ei}}

 \def\abstract#1{\vskip 2 true cm
 	\noindent{\hfill\vbox{\hsize = 14 true cm #1}}\hfill}

\title{The Character of Transport Caused by ExB Drift Turbulence}
\author{Bruce D. Scott}
\address{Max-Planck-IPP, EURATOM Association, 85748 Garching, Germany}
\date{Dec 2001}
 
\abstract{
The basic character of diffusive transport in a magnetised plasma
depends on what kind of transport is modelled.  ExB turbulence under
drift ordering has special characteristics: it is nearly incompressible,
and it cannot lead to magnetic flux diffusion if it is electrostatic.
The ExB velocity is also related to the Poynting energy flux.
Under quasineutral dynamics, electric fields are not caused by transport
of electric charge but by the requirement that the total current is
divergence free.  Consequences for well constructed computational
transport models are discussed in the context of a general mean field
analysis, which also yields several anomalous transfer mechanisms not
normally considered by current models.
}

\vfill

\leftline{PACS numbers: 52.25.Fi  91.25.Cw  52.30.-q  52.40.Nk}

\par\eject

\section{I. Introduction --- Transport as Turbulent Diffusion}

Transport of thermal energy and particles in magnetically confined,
laboratory plasmas is well known to be anomalous: much larger than
transport by collisional diffusion and appearing with different scaling
characteristics
\cite{\hugill,\wootton}.
Modelling of this transport is usually done in terms 
of the two fluid Braginskii equations 
\cite{\brag},
but by substituting either empirical or theoretical models for the
diffusivity coefficients
\cite{\anommodels}.  
The transport process is
assumed by the models to have the same character as the collisional
diffusion process, including the prospect that Onsager symmetry should
hold
\cite{\boozer}.
It is important to note that the transport process is assumed by such a
model to have the same character as the collisional diffusion process.
Many of the results of kinetic transport theory rest on the assumption
that fluctuations in the thermodynamic state variables are small enough
to be neglected in the energy budget and that they are randomly
correlated with each other.

Small scale, low frequency drift
turbulence involving eddies, waves, or vortices of the ExB
velocity,
$$\vexb = {c\over B^2}\vec B\cross\grad\phi,
	\eqno\eqn$$
where at drift scales the perpendicular electric field is 
$\Eperp = -\grad\phi$, 
is a fundamentally different process from this.  Rather than involving
dissipation directly, the ExB eddies advect the background thermal
gradient to produce disturbances in all the thermodynamic state
variables.  The disturbances are then carried with the flows for as long
as the eddies last.  Then, they are picked up by new eddies and carried
further.  This does have the nature of a diffusive process if the size
of the eddies and their lifetime are small and short compared to the
scales of the background, and the net time-averaged transport does
indeed scale with the background gradients if the turbulence is local
\cite{\dalfloc}.  Moreover, although the dynamics of the turbulence is
robustly nonlinear, the relative amplitudes of the disturbances are
indeed small because the plasma is magnetised; the relative amplitudes
$e\phifl/T_e$ or $\pefl/p_e$, for the electrostatic potential and
electron pressure for example, need only be so large as the ratio 
of the local drift scale, $\rs$, to the background gradient scale
length, $L_p$, both defined by
$$\rs^2={c^2 M_iT_e\over e^2B^2} \qquad\qquad
	L_p = \abs{\grad\log p_e}^{-1}
	\eqno\eqn$$
in order to be so nonlinear.  Typical measured values of these
fluctuation amplitudes are of order one to ten percent, rising towards
the edge of a confined plasma due to the fact that $L_p$ becomes much
smaller than the nominal parallel length scale $qR$ (the field line
pitch parameter times the major radius of a toroidal plasma) and the
perpendicular drift dynamics is more able to compete with the
dissipation incurred through the parallel electron dynamics
\cite{\gyro}.  Nevertheless, the strong nature of the coupling
mechanisms between the state variables (especially $\pefl$ and $\phifl$)
means that the basic properties of this turbulence are those of
nonlinear drift waves
\cite{\hasmim,\wakhas},
under drift ordering --- the scale of the turbulence is short compared
to that of the background gradients and the relative amplitude of the
fluctuations in the thermodynamics quantities is small
\cite{\driftorder}.
Drift wave turbulence involves dynamics on scales which are proportional
but not necessarily equal to $\rs$ for space and $c_s/L_p$ for time, for
plasmas which have both the local drift parameter, $\delta$, and the
electron dynamical beta, $\beta_e$, as small parameters.  These two
parameters and the sound speed, $c_s$, are given by
$$\delta={\rs\over L_p} \qquad\qquad
	\beta_e = {4\pi p_e\over B^2} = {c_s^2\over v_A^2}\qquad\qquad
	c_s^2={T_e\over M_i}
	\eqno\eqn$$
where $v_A$ is the Alfv\'en velocity.  The presence of $T_e$ and $M_i$
reflects coupling between electron thermal dynamics and ion inertia.
Even with finite ion temperature $T_i\ne T_e$, the coupling between
the electron pressure and the electric field keeps $\rs$ and $c_s/L_p$
as the typical scaling parameters 
of the disturbances.  This coupling and the fact
that one of the dynamical state variables ($\phifl$)
serves simultaneously as the potential for the electric field and the
stream function for the (essentially incompressible) flow eddies are the
two most important properties of drift wave turbulence.  These
properties and the result that the disturbances are not statistically
independent of each other have consequences concerning the nature of the
transport process.
Our interest in this article is to explore how
these properties differ from those of the kinetic diffusion process
behind classical (and neoclassical) transport processes, and what
implications this has for well constructed transport models.

\section{II. The Basic Character of Small Scale ExB Flow
Transport} 

There are a number of fundamental properties of ExB drift turbulence
which suggest that the transport they cause should be qualitatively
different from random thermal transport.  First, if conventional drift
ordering holds the velocity is nearly incompressible.  If the magnetic
field is straight then the divergence $\div\vexb$ vanishes entirely.  In
a toroidal device, the divergence carries the scale of the major radius:
$$\div\vexb = \div {c\over B^2}\vec B\cross\grad\phi
	\approx -\vexb\cdot\grad\log B^2 \sim R^{-1}\vexb
	\eqno\eqn$$
Since in the confinement zone the pressure scale length, $L_p$ is much
smaller, we have
$$\vedl \pfl \gg \pfl\div\vexb \qquad
	\hbox{if} \quad L_p\ll R
	\eqno\eqn$$
Here, we assume that the scale of a significant divergence of the
transport flux is $L_p$ rather than the much smaller scale of ExB
disturbances, because a locally divergence free state for the fluxes
should be established on the relatively fast time
scale of the turbulence, not the transport.  By contrast, a diffusive
flux of the form $n_e\vv_D = -D\grad n_e$ is always compressible, as
it can be derived from a potential.

The second important characteristic is that the ExB velocity is also
involved in the 
Poynting energy flux,
$$\vexb = {c\over B^2}\vec E\cross\vec B 
	= {1\over \rho v_A^2}\, c {\vec E\cross\vec B\over 4\pi}
	\eqno\eqn$$
with $\rho$ the mass density.
Due to the close force balance for small
scale disturbances in drift dynamics, there is a cancellation
between the Poynting flux and one factor of $p\,\vexb$ in the total
thermal energy flux.
This has the consequence that alone among transport mechanisms, for ExB
drift turbulence the
advective part of the thermal energy flux appears with a factor of 3/2,
not 5/2, times the temperature times the particle flux
\cite{\refthreehalves}.
More
specifically, its energetic coupling to the thermal reservoir is
mediated through its small, quasistatic divergence, so that the pressure
does little work on elemental volumes of the fluid.

A third characteristic of ExB turbulent transport also arises from the
connection between the flow stream function and the electric field:
anomalous energy exchange channels between the electron and ion thermal
energies.  The basic mechanism underlying drift waves is the coupling
between electron pressure and electrostatic disturbances through the
parallel current: $\pefl\fromto\Jfl\fromto\phifl$.  The second step in
this is the Alfv\'enic coupling $\Jfl\fromto\phifl$ which is present in
MHD.  The first step is the adiabatic coupling $\pefl\fromto\Jfl$, part
of which is the adiabatic response in the Ohm's law, by which pressure
disturbances launch shear Alfv\'en waves along the background magnetic
field due to the force balance for electrons parallel to the field.
This adiabatic coupling is responsible for a high degree of correlation
between the electrostatic potential and electron pressure disturbances,
and it provides the means by which the free energy in the background
gradient is passed to the ExB turbulence
\cite{\dalfloc,\wakhas}.  It can only be addressed by a proper two
fluid model which treats electron and ion dynamics separately but self
consistently. 
Since the ExB drift kinetic energy is essentially in the ions, and the
conservation of charge keeps the dynamics quasineutral, a parallel
current divergence in the Alfv\'en dynamics is balanced by a
polarisation current divergence in the ions, and this can do work on the
ion fluid.  Through the disturbances, electron thermal energy taken out
of the background gradient can be deposited in the ions.

A fourth characteristic of ExB flows is that they cannot cause magnetic
flux diffusion unless there is a significant amount of reconnection of
magnetic field lines in the presence of a current gradient (that is, the
drive mechanism should be this current gradient, rather than the thermal
gradients as for drift turbulence).  The magnetic flux occurs through
the inductive part of the electric field in the force balance for
electrons, that is, $\ppt{\Apl}$.  Neglecting dissipation and electron
inertia the Ohm's law is
$${1\over c}\ptt{\Apl} = {\dpl p_e\over n_e e} - \dpl\phi
	\eqno\eqn$$
In an average over a closed flux surface the right hand side vanishes
unless there are appreciable magnetic disturbances.  When the
disturbances follow from drift dynamics the relative phase shifts are
such that magnetic transport effects tend to cancel out of the flux
surface averages.  Only when the current gradient is available as an
energy source do these processes lead to appreciable transport, even if
the thermal transport is robustly anomalous.  By contrast, random
thermal diffusion transports magnetic flux by the same mechanism as for
particles and energy, and the flux diffusivity, $\npl c^2/4\pi$, is even
larger than the fluid diffusivities, proportional to $D_e =
\rho_e^2\nu_e$, by a factor of $\beta_e^{-1}$.  

Two further considerations involve the development of large scale ExB
flows within the flux surfaces (we assume that the equilibrium is
quiescent enough that there is no bulk flow across flux surfaces beyond
the existence of the turbulence).  For dynamics at drift scales, the
disturbances as well as the background are deeply quasineutral.  Charge
differences are neutralised on time and space scales which are
effectively arbitrarily small, so that while there can be nonvanishing
electric fields, there is no significant charge density.  In this
context it is inappropriate to speak of a radial electric field
generated by the divergence of a radial current.  Rather, we have the
constraint that the total current is divergence free, so that various
pieces of the current which might have a divergence are balanced by a
divergence in the polarisation current of the ions.  This is relatively
well known in a general sense, but we will illustrate it here
specifically for the problem of mean flow generation.  If source or loss
mechanisms for either of the charge species are considered, these should
be thought of as a torque on the ExB vorticity rather than a charge
generation\cite{\lackner}.

Finally, anomalous transport in the ExB vorticity results fundamentally
from anomalous viscosity, either the fluid Reynolds stress or some
correction arising from gyroviscosity.  Under flux surface and short
time scale averaging, it is a matter of the total radial polarisation
drift (linear plus nonlinear) necessarily vanishing.  An anomalous
resistivity cannot lead to anomalous transport of ExB flows or,
equivalently, generation of a radial electric field.  This is a direct
consequence of the fact that resistivity is a momentum conserving
friction between electrons and ions.

In the rest of this article, we address the general equations of
transport by small scale drift turbulence via mean field theory, and
then underscore the above points via analysis of these equations.

\section{III. Global Fluid Drift Equations and Energy
Conservation}

The general derivation of fluid drift equations for large scale motion
proceeds similarly to the local treatments more familiar from the
turbulence studies
\cite{\hasmim,\wakhas}.
The the difference is mainly that we must keep
track of the variability of the coefficients which depend on densities
or temperatures.  The density is the most important of these.  It
carries the consequence that the ion polarisation drift must be retained
in advection if exact energy conservation is to hold.  This result
follows from the fact that proper consideration of the electron force
balance (the adiabatic response) disallows the use of the more familiar
MHD ordering in the derivation of global equations
\cite{\hinton,\redmhd,\drake}.
What we must do is
to generalise sufficiently to retain a consistent superset of both
approaches. 

We begin with a neutral, single-component plasma with ions of mass $M_i$
and charge state $Z$ and electrons of mass $m_e$.  For this treatment we
will neglect electron inertia but keep $m_e$ in combination with $\nu_e$
in the formulae for the dissipative fluxes.  The densities satisfy $n_e
= Zn_i$.  We assume that the local drift parameter, $\delta$, and the
electron dynamical beta, $\beta_e$, defined in the Introduction,
are small.  In contrast to
the familiar situation with drift equation models, no assumption
concerning flute mode ordering in the derivatives is made
during the derivation of the equations themselves and their energy
theorem concerning spatial scales, even for $\bb$.  The more familiar
assumption of $\kpl\ll\kpp$, or for a toroidally confined plasma $L_p\ll
R$, is reserved for evaluating the properties of the transport
considering that the disturbances involved in the turbulence do indeed
satisfy local drift ordering.

The drift approximation basically states that perpendicular
compressional wave
dynamics is arbitrarily fast compared to the physics of interest and is
therefore neglected in favor of a quasistatically evolving force balance
perpendicular to the magnetic field lines.  We therefore have a
quasistatic perpendicular electric field, with magnetic induction
appearing only in the parallel component,
$$\Eperp = -\dpp\phi \qquad\qquad 
	\Epl = -{1\over c}\ptt{\Apl} - \dpl\phi
	\eqno\eqn$$
Correspondingly, the total magnetic field, $\bb_t$, is given by the
background, plus the principal disturbance arising from $\Apl$, plus a
small compressional piece which evolves quasistatically, plus an even
smaller correction needed to ensure that $\div\bb_t=0$,
$$\bb_t = \bb - {\bb\over B}\cross\grad\Apl 
	+ \Bfl\,{\bb\over B} - \dpp\chi_B
	\eqno\eqname\eqbb$$
respectively.  The symbols $\bb$ and $B$ give the equilibrium magnetic
field and its magnitude, while $\bunit$ is reserved for the combination
of $\bb$ and the piece involving $\Apl$, given by
$$\Bperp = - {\bb\over B}\cross\grad\Apl 
	\eqno\eqn$$
so that in describing parallel components we use
$$\bunit = {1\over B}\LP\bb+\Bperp\RP \qquad\qquad
	\dpl = \bdel
	\eqno\eqn$$
Perpendicular components are described by the background field only,
$$\dpp = - \bb\cross{\bb\cross\grad\over B^2} \qquad\qquad
	\ddpp = \div\dpp
	\eqno\eqn$$
so we must note that with these definitions we cannot write
$\grad=\dpp+\bunit\dpl$, for example.
The Ampere's law involving $\Apl$ is written with $\ddpp$,
$$ \ddpp\Apl = -{4\pi\over c}\Jpl
	\eqno\eqname\eqampere$$
The compressional part of the magnetic field, involving $\Bfl$, is
written for generality but not used in a $\beta_e\ll 1$ treatment (it
can be used to evaluate the Poynting energy flux in the energy theorem
by assuming that $\ddpp$ of the sum $B^2+8\pi p$ is zero).  The
divergence correction, involving $\chi_B$, is never used; it is written
only to preserve $\div\bb_t=0$, and in practice it is always negligibly
small. 

The quasistatic force balance allows us to write the 
drift velocities and heat fluxes
perpendicular to $\bb$ in terms of the state variables.  For electrons
the diamagnetic fluxes are
$$\vstar = - {1\over n_e e}\drift\grad p_e \qquad
	\qestar = - \fivehalves{p_e\over e}\drift\grad T_e 
	\eqno\eqname\eqdrifts$$
and for ions
$$\ustar = {1\over n_i Ze}\drift\grad p_i \qquad
	\qistar = \fivehalves{p_i\over Ze}\drift\grad T_i 
	\eqno\eqname\eqidrifts$$
The ExB drift for both species and the diamagnetic current are given by
$$\vexb=\drift\grad\phi \qquad\qquad
	\Jstar = n_e e (\ustar-\vstar) = \drift\grad p
	\eqno\eqname\eqexbdrift$$
with $p=p_e+p_i$.  We form the lowest order perpendicular fluid
velocities resulting from these drifts, 
$$\uperp = \vexb + \ustar \qquad\qquad
  \vperp = \vexb + \vstar 
	\eqno\eqn$$
In the parallel dynamics the dependent variables are $\upl$ and $\Jpl$,
with $\vpl$ defined as
$$\vpl = \upl - {\Jpl\over n_e e}
	\eqno\eqname\eqvpl$$
The parallel current is actually an auxiliary variable obtained from 
$\Apl$ through the drift form of Ampere's law, Eq.~(\eqampere) above,
with $\delsq$ replaced by $\ddpp$.

For the ions we keep the polarisation drift, $\upol$, which is required
in the current balances and also in the contributions of the ion
pressure to energy conservation.  The polarisation drift is obtained by
inserting the lowest order ion velocity, $\uperp+\upl\bunit$, into the
ion inertia terms (including all components of the stress tensor), and
solving anew for the total ion velocity,
$$\uu=\uperp+\upl\bunit+\upol
	\eqno\eqname\eqivelocity$$
For the electrons we neglect polarisation along with $m_e$, so that
their total velocity is
$$\vv=\vperp+\vpl\bunit
	\eqno\eqname\eqevelocity$$
noting that $\vpl$ is given in Eq.~(\eqvpl).
The polarisation drift is given by
$$n_e e \upol = n_i M_i \drift \LP\ptt{}+\udel\RP 
	\LB\drift\LP \grad\phi+{\grad p_i\over n_e e}\RP\RB
	+ \drift(\div\pistar)
	\eqno\eqn$$
where we keep the gyroviscosity explicitly instead of just using the
part which cancels $\ustar\cdot\grad$, in order to preserve its
symmetry.  We can express the polarisation drift in terms of a
potential,
$$\upol = -{M_i c^2\over Z eB^2}\dpp\chi 
	= {M_i\over Ze}\,\drift\LP\drift\grad \chi\RP
	\eqno\eqname\equpol$$
following from the fact that it is primarily proportional to gradients
of scalars.  In contrast to the local treatments, the polarisation drift
does indeed appear explicitly in the inertia, but only in the advective
derivative.  On the other hand, the inertial velocity itself is the
lowest order version, whose perpendicular component is $\uperp$.
Consequently, we construct the components of $\pistar$ using only the
lowest order velocities, $\uperp+\upl\bunit$, as we will need to use the
fact that $\pistar\dotdot\grad\uu$ vanishes, for whichever vector $\uu$
is used to construct $\pistar$ (if this is done in detail it must be
extended to include $\qistar$ in $\pistar$ to get the gyroviscous
cancellation
\cite{\hinton}
to work properly when the temperature and density gradients are equally
steep
\cite{\smolyakov}).
We note here that everything having to
do with ion inertia in the perpendicular force balance is encompassed
within the polarisation drift, including not only the advection of
vorticity but also Reynolds stress and gyroviscosity.

The equations under the drift approximation are formed by substituting
the above expressions for the velocities into the rest of the fluid
equations.  The evolution of $\phi$ is controlled by the polarisation
equation, which is formed by subtracting the densities so as to form a
continuity equation for electric charge.  The charge density itself is
set to zero, leaving $\div\jj=0$.  Expressing the total current as
$$\jj=\Jpl\bunit+\Jstar+\Jpol
	\eqno\eqn$$
with the polarisation current given by $\Jpol=n_e e\upol$, we find
$$-\div n_e e\upol 
	= \div\LP\Jpl\bunit\RP + \div\drift\grad\LP p_e+p_i\RP
	\eqno\eqn$$
Where we need to evaluate $\upol$ in the ion equations, we may use,
$$\div {n_i M_i c^2\over B^2}\dpp\chi
	= \div\LP\Jpl\bunit\RP + \div\drift\grad\LP p_e+p_i\RP
	\eqno\eqname\eqchi$$
which we solve for $\chi$.  The evolution of $\Apl$ and $\Jpl$ is
controlled by the Ohm's law, which is just the parallel electron force
balance including dissipation.  The rest of the dependent variables in
the set
$\{\phi,n_e,\Apl,p_e,\upl,p_i\}$ have their own fluid equations, in
which we use the same procedure as in the polarisation equation.
The resulting model is given by
$$-\div \LBR n_i M_i \drift \LP\ptt{}+\udel\RP 
	\LB \drift\LP \grad\phi+{\grad p_i\over n_e e}\RP \RB
	+ \drift(\div\pistar) \RBR
={}\hskip 1 cm$$$$\hskip 7 cm{}
	= \div \LP\Jpl\bunit\RP + \div\drift\grad\LP p_e+p_i\RP
	\eqno\eqname\eqvor$$
$$\ptt{n_e} + \div \LB n_e\vexb 
	+ n_e \LP\upl-{\Jpl\over n_e e}\RP\bunit 
	- \drift\grad {p_e\over e} \RB = 0
	\eqno\eqname\eqne$$
$${1\over c}\ptt{\Apl} = {\dpl p_e\over n_e e}-\dpl\phi 
	-R_{ei}
	\eqno\eqname\eqpsi$$
$$\ptt{}\threehalves p_e + \div \LB \threehalves p_e\vexb
	+ \fivehalves p_e\LP\upl-{\Jpl\over n_e e}\RP\bunit 
	+ \qepl\bunit
	- \drift\grad\fivehalves{p_e T_e\over e} \RB
={}\hskip 2 cm$$$$\hskip 6 cm{}
	= \upl\dpl p_e - {\Jpl\over n_e e}\dpl p_e - p_e\div\vexb
	+ \Jpl R_{ei}
	\eqno\eqname\eqpe$$
$$n_iM_i\LP\ptt{}+\udel\RP\upl + \bunit\cdot(\div\pistar)
	= -\dpl\LP p_e+p_i\RP + \div\LP\mupl\bunit\dpl\upl\RP
	\eqno\eqname\equi$$
$$\ptt{}\threehalves p_i + \div \LB \threehalves p_i\vexb
	+ \fivehalves p_i\upl\bunit 
	+ \qipl\bunit
	+ \fivehalves p_i\upol 
	+ \drift\grad\fivehalves{p_i T_i\over e} \RB
={}\hskip 2 cm$$$$\hskip 6 cm{}
	= \upl\dpl p_i + \upol\cdot\grad p_i - p_i\div\vexb 
	+ \mupl\abs{\dpl\upl}^2
	\eqno\eqname\eqpi$$
where except for $\div\vexb$ the magnetic divergences are written
explicitly.  The dissipative momentum transfer,
$$R_{ei} = \npl\LB\Jpl+{\alpha\over\kappa_e}
		\LP\alpha\Jpl+{e\over T_e}\qepl\RP\RB
	\eqno\eqname\eqrei$$
and parallel heat fluxes,
$$\qepl + \alpha {T_e\over e}\Jpl 
	= -\kappa_e n_e {V_e^2\over\nu_e} \dpl T_e \qquad\qquad
\qipl = -\kappa_i n_i {V_i^2\over\nu_i} \dpl T_i
	\eqno\eqn$$ 
are given by the Braginskii model,
with the resistivity and electron and ion thermal velocities given by
$$\npl = \eta{m_e\nu_e\over n_e e^2} \qquad\qquad
V_e = \LP{T_e\over m_e}\RP^{1/2} \qquad\qquad
	V_i = \LP{T_i\over M_i}\RP^{1/2}
	\eqno\eqn$$
respectively, with numerical constants for $Z=1$
$$\eta = 0.51 \qquad\qquad \alpha=0.71 \qquad\qquad
  \kappa_e = 3.2 \qquad\qquad \kappa_i = 3.9
	\eqno\eqn$$
and the collision frequencies given by the inverses of the Braginskii
collision times
\cite{\brag}.

We note that this is an approximate treatment, but as we will now show,
it conserves global energy.  We first make a few remarks.  First, the
reason for including $\upol$ in the advective derivative $\udel$ is
conservation of the inertial energy incorporated in $\uperp$.  We must
manipulate the polarisation equation such that the density is brought
under both the time derivative and under a resulting total divergence.
Since it starts out as a multiplier on $\ppt{}$ as well as on the
advecting velocity, the continuity demands require that we keep $\upol$
in $n_i\udel$ if we keep it in $\div n_i\uu$.  This chain of demands
starts at the presence of $\dpl p_e$ in the Ohm's law, continues with
$\dpl\Jpl$ in the electron pressure equation and hence the density
equation, and finally with the fact that the divergences of $\Jpl$ and
$\upol$ go together in the continuity.  The ordering and conservation
are preserved by keeping $\upol$ in $\udel$, but not in the actual
inertial velocity, $\uperp$.  Second, although the perpendicular and
parallel components of vectors are set up differently, they are kept
separate in all computations so that there is no violation of energy
conservation.  Finally, one sees that part of the drift ordering is
expressed in the fact that the equations are written only for the lowest
order components of the ion velocity and magnetic field, but the
operations involving those velocities and gradients include all
components in order to conserve energy.

The energy theorem is formed by multiplying Eqs.~(\eqvor,\equi) for the
vorticity and parallel velocity by $-\phi$ and $\upl$, and Eq.~(\eqpsi)
for the magnetic potential by $\Jpl$, respectively.  The pressures in
Eqs.~(\eqpe,\eqpi) represent the thermal energy.  The total energy
density is
$$U = \int d^3 x\LP 
	n_i M_i {\abs{\uperp}^2\over 2} 
	+ n_i M_i {\upl^2\over 2} 
	+ \threehalves p_e + \threehalves p_i
	+ {\abs{\Bperp}^2\over 8\pi} \RP
	\eqno\eqname\eqenergydef$$
We will call these pieces the drift energy, sound wave energy, electron
and ion thermal energy, and magnetic energy, respectively.

The first consideration is how polarisation works under the drift
approximation.  
We multiply the polarisation equation by $\phi$, to obtain
$$\phi\div\jj = \div\phi\jj - \jj\cdot\grad\phi = 0
	\eqno\eqn$$
where $\jj$ includes all three pieces, $\Jpl\bunit+\Jstar+\Jpol$.  The
form of this that we actually use is
$$-\Jpol\cdot\grad\phi + \div\phi\jj - \Jstar\cdot\grad\phi
	= \Jpl\dpl\phi
	\eqno\eqn$$
where the first piece will give part of the drift energy, the next two
pieces will be combined into a total flux plus a transfer term, and the
piece on the right side is already a transfer term.  We manipulate
$$\Jstar\cdot\grad\phi = \LP\drift\grad p\RP\cdot\grad\phi
	= - \LP\drift\grad\phi\RP\cdot\grad p = -\vedl p
	\eqno\eqn$$
where $p=p_e+p_i$ is the total pressure,
and then combine with the total divergence to write
$$-\Jpol\cdot\grad\phi + \div\LP\phi\jj + p\vexb\RP
	= \Jpl\dpl\phi + p\div\vexb
	\eqno\eqname\eqdriftena$$
We note that in general the divergence of $\phi\jj$ is
equivalent to part of the Poynting flux,
$$\div\phi\jj = \div\LP{c\over 4\pi}\phi\curl\bb_t\RP
	= - \div\LP{c\over 4\pi}\grad\phi\cross\bb_t\RP
	\eqno\eqn$$
dropping a divergence of a curl.  The
near cancellation with $p\vexb$ is called the Poynting cancellation,
$$\div\LP\phi\jperp + p\vexb\RP\approx 0
	\eqno\eqname\eqpoyntingcancel$$
affecting $\Jperp$ since $\vexb$ is also perpendicular to $\bb$.  
This cancellation results in the fact that the ExB thermal transport
eventually appears with a factor of $3/2$, so that the ExB divergence,
not an ExB advection, is the process by which transfer with the thermal
reservoir occurs.  We will show this in Section V, together with the
other characteristics of ExB transport.

The part of the Poynting flux involving $\Jpl$ moves parallel to $\bb$
and is the process by which the MHD part of shear Alfv\'en waves move
field energy along $\bb$:
$$\div(\phi\Jpl\bunit)\approx
	- \div\LP{c\over 4\pi}\grad\phi\cross\Bperp\RP
	\eqno\eqname\eqalfvenpoynting$$
neglecting the action of the curl on $\bunit$ itself.  This is generally
comparable to or larger than $\div(p_e\vpl\bunit)$, reflecting the
properties of shear Alfv\'en dynamics under general two fluid ordering.

The pressure piece in the drift energy
is found by subtracting $\upol\cdot\grad p_i$ from both
sides of Eq.~(\eqdriftena), to find
$$-\Jpol\cdot\LP\grad\phi+{\grad p_i\over n_e e}\RP
	+ \div\LP\phi\jj + p\vexb\RP
	= \Jpl\dpl\phi + \LP p_e+p_i\RP\div\vexb - \upol\cdot\grad p_i
	\eqno\eqn$$
Finally, substituting in for $\Jpol$ and noting how the combination in
parentheses gives rise to $\uperp$, we have
$$\ptt{}\LP n_iM_i{\abs{\uperp}^2\over 2}\RP
	+ \div \LB n_i M_i{\abs{\uperp}^2\over 2}\uu
	+ \uperp\cdot\pistar
	+ \phi\jj + p\vexb \RB
={}\hskip 1 cm$$$$\hskip 4 cm{}
	= \pistar\dotdot\grad\uperp + 
	\Jpl\dpl\phi + \LP p_e+p_i\RP\div\vexb - \upol\cdot\grad p_i
	\eqno\eqname\eqendrift$$
in which the transport terms are the ones under the divergence operator
on the left side, and the transfer terms are those on the right side.
We note that the reason that $\uu$ must include all components of the
ion velocity is to bring the factor of $n_i$ under both the partial time
derivative and the divergence operator simultaneously.  In other words,
whatever we keep in the velocity divergence in the density equation must
also be kept in the advection in the polarisation equation, and in all
the ion fluid equations as well.

We now find where the energy transferred out of this drift energy goes.
The next subtlety is the magnetic energy.  We 
multiply Eq.~(\eqpsi) by $\Jpl$ and manipulate the divergence to obtain
$$\ptt{}{1\over 8\pi}\abs{\dpp\Apl}^2
	+ \div\LP\dpp\Apl{1\over 4\pi c}\ptt{\Apl}\RP
	= {\Jpl\over n_e e}\dpl p_e - \Jpl\dpl\phi
	-\Jpl R_{ei}
	\eqno\eqn$$
The terms on the right side are the adiabatic and
Alfv\'enic transfer effects and
resistive dissipation, which function the same way as in
drift wave turbulence
\cite{\dalfloc}.
The divergence term is the inductive part of the Poynting energy flux,
as we can see by evaluating the projection operations implicit in the
$\dpp$ operator,
$$\div\LP\dpp\Apl{1\over 4\pi c}\ptt{\Apl}\RP
	= \div {c\over 4\pi}\LB{1\over c}{\bb\over B}\ptt{\Apl}\cross
		\LP{\bb\over B}\cross\grad\Apl\RP\RB
	\eqno\eqn$$
Substituting $\Bperp$ for $\Apl$,
this gives the conservation law for the perturbed magnetic energy
$$\ptt{}{\abs{\Bperp}^2\over 8\pi}
	- \div {c\over 4\pi}\LP {1\over c}\bunit\ptt{\Apl}\cross\Bperp \RP
	= {\Jpl\over n_e e}\dpl p_e - \Jpl\dpl\phi 
	-\Jpl R_{ei}
	\eqno\eqname\eqenmag
$$
where we note that the inductive electric field appears with $\bunit$
since the term it is crossed into is $\Bperp$.  The inductive part of
the Poynting flux involving $\Bperp$ is in this equation, transporting
magnetic energy, while the static part is in Eq.~(\eqalfvenpoynting),
transporting ExB energy.  Both are small, however, except for the part
propagating along $\bb$, which carries the Alfv\'en wave energy but
whose contribution to transport across magnetic flux surfaces is
nevertheless negligible. 

The sound wave energy is found by multiplying Eq.~(\equi) by
$\upl$, noting that here also the need to put the factor of $n_i$ under
the time derivative and the divergence requires keeping $\upol$ in the
advection terms.  We obtain
$$\ptt{}\LP n_i M_i{\upl^2\over 2}\RP
	+ \div\LP n_i M_i{\upl^2\over 2}\uu + \upl\bunit\cdot\pistar
	- \upl\bunit\,\mupl\dpl\upl \RP
={}\hskip 1 cm$$$$\hskip 4 cm{}
	= \pistar\dotdot\grad\LP\upl\bunit\RP
	- \upl\dpl \LP p_e+p_i\RP
	- \mupl\abs{\dpl\upl}^2
	\eqno\eqname\eqensound$$
noting that the gyroviscosity term cancels properly with the one in
Eq.~(\eqendrift) because $\pistar\dotdot\grad(\uperp+\upl\bunit)$ vanishes.
Gyroviscosity thereby represents a diamagnetic momentum flux which acts
to transfer energy between parallel and perpendicular fluid motion.

The thermal energy comes next.  These are simply the pressure equations,
with some of the advection terms evaluated piece by piece.  For the
electrons we have Eq.~(\eqpe), in which the transfer effects with other
equations are now obvious: adiabatic transfer ($\Jpl\dpl p_e$) and
resistive frictional heating $(\Jpl R_{ei})$ with the magnetic energy,
acoustic coupling ($\upl\dpl p_e$) with the sound waves, and the ExB
divergence ($p_e\div\vexb$) with the drift energy through polarisation.
For the ions we have Eq.~(\eqpi), in which the transfer effects are
thermal coupling ($\upl\dpl p_i$) and viscous heating (the $\mupl$ term)
with the sound waves, and the ExB divergence ($p_i\div\vexb$) through
polarisation and the polarisation advection ($\upol\cdot\grad p_i$) with
the drift energy.

The total energy theorem is then given by
$$\ptt{U} + \div\bigg[
	\threehalves (p_e+p_i)\vexb + \LP\qepl+\qipl\RP\bunit
+{}\hskip 6 cm$$$$\hskip 0.5 cm{}
	+ \fivehalves\LP \phi\Jpl+p_e\vpl+p_i\upl \RP\bunit
	+ \fivehalves p_i\upol
	+ \drift\grad\fivehalves{p_i T_i\over e}
	- \drift\grad\fivehalves{p_e T_e\over e} 
+{}\hskip 0.5 cm$$$${}
	+  n_i M_i{\abs{\uperp}^2+\upl^2\over 2}\uu
	+ \pistar\cdot\LP\uperp+\upl\bunit\RP
	+ \LP\phi\jperp + p\vexb\RP
	- {c\over 4\pi} {1\over c}\bunit\ptt{\Apl}\cross\Bperp
	\bigg] = 0
	\eqno\eqname\eqtotalenergy$$
The transport effects are written in order of their usual importance; in
practical situations only the terms in the first line need be kept.  In
situations where transport by Pfirsch-Schl\"uter currents (generally
caused by nonzero magnetic divergences involving the equilibrium
pressure) is relevant, the second line would have to be kept as a unit,
since the divergences of all the currents go together.  The diamagnetic
fluxes are combined into the terms involving the drift operator.  The
terms in the last line are negligible unless there are transonic or
supersonic flows.

The dominant transport effect across the magnetic flux surfaces is the
ExB advection, $(3/2)p_e\vexb$, which appears with the factor of $3/2$
due to the Poynting cancellation, Eq.~(\eqpoyntingcancel).  The
diamagnetic fluxes drop to the second line due to the well known
diamagnetic cancellation in the pressure equations 
\cite{\tsai}.
For the
parallel transport, the principal effects are the electron and ion heat
fluxes, $B^{-1}(\qepl+\qipl)\Bperp$.  We also have the ion advection,
$(5/2B)p_i\upl\Bperp$, appearing with the factor of $5/2$ due to the
compressibility of the parallel flows.  For the electrons the part of
this effect involving $\Jpl$ is
small because $\Apl$ is the stream function for $\Bperp$ and it is also
related to $\Jpl$ through Ampere's law, but here we also have the
acoustic advective effect, $(5/2B)p_e\upl\Bperp$.  These parallel
advective effects are small, however, since the sound wave transit time
is generally much longer than the time scale of the turbulence.

To summarise, we have a set of drift equations following the drift
ordering as to magnetic compression but allowing for an arbitrary scale
of motion and with it, interactions with the thermal gradient of the
background.  The conservation of total energy is exact, but at the price
that the total velocity must be kept in advection.  The extra piece,
that is, the polarisation drift, is given in terms of the dependent
variables through a constitutive relation, Eq.~(\eqchi).

\section{IV. A One Dimensional Mean Field Model 
of Transport by ExB Turbulence}


In this section we construct a mean field transport model including both
perpendicular and parallel flows.  All dependent variables are split
into profile quantities which are flux functions (dependence only on the
flux surface label coordinate), and disturbances which have arbitrary
coordinate dependence but are zero in ensemble average.  The profiles
are assumed to be slowly varying in time and space compared to the
disturbances, so the ensemble average can be understood as one over an
interval of time short compared to the transport time of a given radial
region but long compared to the fluctuation time scale of the
turbulence.  The energy content of the disturbances is assumed to be
negligible, so that their energy equations become statements that their
net transfer effects are all in quasistatic balance.  Transport of wave
energy across flux surfaces (disturbance energy content times $\vefl$)
is kept until the 
disturbance energy equations are evaluated, in order to show that it is
small. The disturbances
form nonvanishing quadratic transport and transfer quantities in the
mean field equations; these quantities control the evolution of the
profiles.  

We retain both poloidal and toroidal rotation, fully described by the
two state variables, $\phi$ and $p_i$, and the parallel flow $\upl$.  In
contrast with more usual treatments, we keep the partial time
derivatives in the polarisation equation, and hence also the energy
content of the drift flows.  Both flows are affected by Reynolds stress
and by neoclassical friction, the latter acting only on the poloidal
component of the total flow (recall that parallel is not purely toroidal
and perpendicular is not purely poloidal).  The parallel flow
disturbances are also dissipated by parallel viscosity.  We also retain
magnetic flux diffusion, keeping resistive dissipation for both profiles
and disturbances.

The equations for the profiles are those derived in Section III,
simplified by the symmetry assumptions, but with the addition of the
ensemble averaged transport effects.  For the disturbances we mainly
have to declare what is kept in order to satisfy energy conservation,
since the role of the disturbances is to provide transfer channels.

To proceed with the transfer evaluations we need the density equation.
This model follows transport of electrons, but the ion density is needed
for the factors of $n_iM_i$ in front of the partial time derivatives in
the flow equations to conserve energy properly.  For the electrons we
have
$$\ptt{n_e} + \div\avg{\nefl\vefl} = 0
	\eqno\eqname\eqnemean$$
neglecting compression effects.  With this
simplification, we have equivalently the same equation for the ions
$$\ptt{n_i} + \div\avg{\nifl\vefl} = 0
	\eqno\eqname\eqcontinuity$$
We write both these equations to emphasise the fact that neglect of the
polarisation compression and the neglect of particle transport by
the parallel current in general go together. 
It is important here to note the assumption that $\div n_i\vefl$ can be
neglected, since the density-weighted average of $\vefl$ should vanish
in order to have transport not go by direct flow but by statistical
average.  This helps in the manipulation of the Reynolds stresses.  

The parallel flow profile equation is
$$n_i M_i\ptt{\upl} + M_i\avg{\nifl\vefl}\cdot\grad\upl
	+ \div n_i M_i\avg{\ufl\vefl} 
	+ n_i M_i \nu_d u^\theta b^\theta
	= 0
	\eqno\eqname\equimean$$
in which we have used the ability to neglect $\div n_i\vefl$ before
taking the ensemble average.
The last term is the parallel component of the poloidal friction, with
$u^\theta=\LP\uperp+\upl\bunit\RP\cdot\grad\theta$ the total poloidal
flow component, $b^\theta=\bdel\theta$ the poloidal component of the
magnetic 
unit vector, $\nu_d$ the neoclassical friction coefficient, and
$\uperp$ given by the profiles of $\phi$ and $p_i$ as in Section III.
The third term is the ``conductive'' parallel momentum transport, as we
can see by using Eq.~(\eqcontinuity) to form the parallel momentum
equation,
$$\ptt{}\LP n_iM_i\upl\RP 
	+ \div\Big( M_i\avg{\nifl\vefl}\upl 
		+ n_i M_i\avg{\ufl\vefl}\Big)
	+ n_i M_i \nu_d u^\theta b^\theta
	= 0
	\eqno\eqname\eqmomentum$$
The first term under the divergence 
is transport via the anomalous particle flux, and the 
second is due to correlations between the parallel flow and ExB
disturbances, which is also the perp-parallel Reynolds stress.
The latter term is called the ``conductive'' part of the transport.
The transport operating through the particle flux is
called the ``convective'' part.  We simply generalise this concept,
which is standard for thermal energy, to the other conserved quantities.
The last term transfers momentum with the magnetic field; 
the momentum of $\uperp$
is not followed, consistent with the drift approximation in which the
background magnetic field is an anchor.

For the parallel flow disturbances we have
$$n_i M_i\ptt{\ufl} + n_i M_i\vefl\cdot\grad\ufl
	+ M_i\avg{\nifl\vefl}\cdot\grad\ufl
	+ n_i M_i\vefl\cdot\grad\upl
	= -\dpl\pfl
	+ \mupl\ddpl\ufl
	\eqno\eqn$$
where $\pfl=\pefl+\pifl$ is the total pressure disturbance.
The second and third terms are nonlinear advection, which is required in
light of Eq.~(\eqcontinuity) by the need to
get the factor of $n_i M_i$ under the time derivative.
The fourth term
is the drive term arising from the background flow gradient.  It
conserves energy against the perp-parallel Reynolds stress in the
profile equation.  The energy equations are given by
$$\eqalign{\ptt{}\LP n_i M_i{\upl^2\over 2}\RP
 &
	+ \div\LP M_i\avg{\nifl\vefl}{\upl^2\over 2}
		+ n_i M_i\avg{\ufl\vefl}\upl \RP
={}\cr&\qquad\qquad{}
	= n_i M_i\avg{\ufl\vefl}\cdot\grad\upl
		- n_i M_i \nu_d u^\theta \upl b^\theta
\cr}
	\eqno\eqname\equienergy$$
for the profile, and 
$$\eqalign{\ptt{}\LP n_i M_i\avg{{\ufl^2\over 2}}\RP
 &
	+ \div\LP n_i M_i\avg{\vefl{\ufl^2\over 2}}
		+ M_i\avg{\nifl\vefl}{\ufl^2\over 2}\RP
	- \dpl\Big(\mupl\avg{\ufl\dpl\ufl}\Big)
={}\cr&\qquad{}
	= - n_i M_i\avg{\ufl\vefl}\cdot\grad\upl
		- \avg{\ufl\dpl\pfl}
		- \mupl\avg{\abs{\dpl\ufl}^2}
\cr}
	\eqno\eqname\equflenergy$$
for the disturbances, in both of which Eq.~(\eqcontinuity) times the
relevant factors of squared velocity (specific energy) is used.  This
also uses $\div n_i\vexb=0$ to treat the conventional nonlinearity, and
in this and the other disturbance energy equations the resulting term in
the energy flux is a third order nonlinearity which represents the ExB
transport of wave energy in the turbulence.  We carry these terms here
for illustration, but since they are down by two orders of $\delta$ they
are always small in practice (and are measured to be small in
turbulence computations). 

The energy transfer via the Reynolds stress is apparent in
Eq.~(\equflenergy), as are the 
parallel flow transfer effects we saw in Section III.  
These four Eqs.~(\equimean,\eqmomentum,\equienergy,\equflenergy)
together show why the particle flux has to be kept in advection;
otherwise we would have problems with simultaneous conservation of
energy and momentum.  The same is true for the perpendicular flow, which
we treat next.

The perpendicular flow is determined by the scalar fields $\phi$ and
$p_i$, and since the ion pressure has its own equation this one becomes
the equation for $\phi$.  For the profiles we have
$$\eqalign{\div {n_i M_i c^2\over B^2} 
	\ptt{}\LP\dpp\phi+{\dpp p_i\over n_e e}\RP
={}\qquad&\cr{}=
	\div\drift\Big( M_i\avg{\nifl\vefl}\cdot\grad\uperp
&
		+ \div n_i M_i \avg{\vefl\vefl}
		+ n_i M_i\nu_d u^\theta\grad\theta\Big)
\cr}
	\eqno\eqname\eqphiprof$$
in which the only effects are nonlinear forcing and frictional damping,
as with the $\upl$ profile.  As in Eq.~(\equimean), we rewrite the
$\vefl\cdot\grad\vefl$ term as a Reynolds stress.
The more familiar linear terms and the conventional vorticity
nonlinearity are all in
the equation for the disturbances, 
$$\eqalign{\div & {n_i M_i c^2\over B^2}\ptt{}\dpp\phifl = {}
\cr & \qquad{}=
	\div \drift \Big( n_i M_i \vedl\vefl
		+ n_i M_i \vefl\cdot\grad\vefl
		+ n_i M_i \vefl\cdot\grad\uperp\Big)
\cr & \qquad\qquad{}
	+ \div\LP\Jfl\bunit\RP + \div\drift\grad\pfl
\cr}
	\eqno\eqname\eqphifl$$
neglecting the frictional damping due to its slow time scale.  
We are keeping only $\phifl$ in the vorticity disturbances, as the ExB
contributions to the Reynolds stress are dominant.  The first term on
the right side illustrates ExB shearing but will drop out of the energy
equation due to symmetry.  The next term is the conventional
nonlinearity and will give rise to the ExB wave energy transport.  The
third term is the advection of the flow profile by flow disturbances,
and is the term which conserves energy against the Reynolds stress in
Eq.~(\eqphiprof). 

The energy equation for the flow profile is constructed as in Section
III: we multiply Eq.~(\eqphiprof) by $-\phi$ and manipulate the
divergences (equivalent to integrations by parts under an integral),
also subtracting $(\Jpol/n_e e)\cdot\grad p_i$ from both sides, the left
side completing the specific energy (squared velocity) and the right
side providing the polarisation transfer with the ion pressure.  
Eq.~(\eqcontinuity) is used to bring the factors of $n_i$ under the
partial time derivatives.
For the profile we have
$$\eqalign{\ptt{}\LP n_i M_i {\upp^2\over 2}\RP &
	+ \div\LP \phi\Jpol + M_i\avg{\nifl\vefl}{\upp^2\over 2}
		+ n_i M_i\avg{\vefl\vefl}\cdot\uperp\RP
={}\cr&\qquad\qquad{}
	= n_i M_i\avg{\vefl\vefl}\dotdot\grad\uperp
		- n_i M_i\nu_d u^\theta\upp^\theta
		- \Jpol\cdot{\grad p_i\over n_e e}
\cr}
	\eqno\eqname\eqphienergy$$
where $\upp^\theta=\uperp\cdot\grad\theta$, and
$$\eqalign{\Jpol & = - n_iM_i\ptt{}\LP\dpp\phi+{\dpp p_i\over n_e e}\RP
+{}\cr&\qquad\qquad{}
	+ \drift\Big(
		M_i\avg{\nifl\vefl}\cdot\uperp
		+ \div n_iM_i\avg{\vefl\vefl}
		+ n_iM_i\nu_d u^\theta\grad\theta\Big)
\cr}
	\eqno\eqname\eqjpol$$
is the polarisation current,
and for the disturbances we have
$$\eqalign{
\ptt{}\LP n_i M_i \avg{{\uppfl^2\over 2}}\RP 
	+ \div& \Bigg( M_i\avg{\nifl\vefl}\avg{{\uppfl^2\over 2}}
		+ n_i M_i\avg{{\uppfl^2\over 2}\vefl}
+{}\cr & \qquad\qquad{}
		+ \avg{\phifl\ptb\Jpol}
		+ \avg{\phifl\ptb\Jpl}
		+ \avg{\phifl\ptb\Jstar}
		+ \avg{\pfl\vefl}\Bigg)
\cr&\qquad{}
	= - n_i M_i\avg{\vefl\vefl}\dotdot\grad\uperp
		+ \avg{\Jfl\dpl\phifl}
		- \avg{\pfl\div\vefl}
\cr}
	\eqno\eqname\eqphiflenergy$$
where $\uppfl^2=\vefl^2$ since we are keeping only ExB vorticity in the
disturbances. 
We wee that the Reynolds stress transfers within the flow energy between
profile and disturbances, and the disturbances then transfer to the
pressures ($\pfl\div\vefl$) and
the magnetic disturbances ($\Jfl\dpl\phifl$).  The wave energy transport
is small by $O(\delta^2)$ compared to either $\avg{p\vexb}$ or
$\avg{\phi\Jperp}$.  The cross flux surface Alfv\'en wave flux
$\avg{\phifl\Jfl\bbfl}$ is also small by $O(\delta^2)$.

The electron and ion pressure equations are somewhat easier, since they
are already written in conservative form.  We keep the most important
quadratic fluxes, plus all the transfer effects of the disturbances.
For the electrons we have
$$\eqalign{\threehalves\ptt{p_e} + \div &\LP
	\threehalves \avg{\pefl\vefl}
	+ \avg{\qefl\bbfl}
	+ \fivehalves p_e\avg{\ufl\bbfl}
	\RP
={}\cr&\qquad{}
	= \avg{\ufl\dpl\pefl} - {1\over n_e e}\avg{\Jfl\dpl\pefl} 
		- \avg{\pefl\div\vefl}
	+ \npl\Jpl^2
	+ \avg{\Jfl\ptb R_{ei}}
\cr}
	\eqno\eqname\eqpemean$$
neglecting the magnetic flutter transport involving $\Jfl$.  
For the ions we have
$$\eqalign{\threehalves\ptt{p_i} + \div &\LP
	\threehalves\avg{\pifl\vefl}
	+ \avg{\qifl\bbfl}
	+ \fivehalves p_i\avg{\ufl\bbfl}
	\RP
={}\cr&\qquad{}
	= \avg{\ufl\dpl\pifl} - \avg{\pifl\div\vefl} 
		+ \avg{\upfl\cdot\grad\pifl}
+{}\cr&\qquad\qquad{}
	+ n_i M_i \nu_d u^\theta u^\theta 
	+ {\Jpol\over n_e e}\cdot\grad p_i
	+ \mupl\avg{\abs{\dpl\ufl}^2}
\cr}
	\eqno\eqname\eqpimean$$
correspondingly neglecting the transport due to $\upol$.  This is the
form of the pressure equations including magnetic flutter transport;
formally the part of that due to $\ufl$ goes together with the transfers
adding up to $\avg{\ufl\dpl\pfl}$, but while the transfer effects
remain, the transport effects are overshadowed by the contributions to
$\avg{\pfl\vefl}$, so we drop flutter transport at this stage.  Since
Reynolds stress is not a proper viscosity, it does not {\it a priori}
appear in the ion pressure equation.  It represents a transfer effect
with the disturbances, which in turn transfer the energy via other
channels with the ion pressure, most notably through sound waves and
through some of the electron channels, as we will see.

The magnetic field evolves through flux diffusion,
$${1\over c}\ptt{\Apl} = E_L - \npl\Jpl
	\eqno\eqn$$
balanced by the loop voltage $E_L$, here also neglecting magnetic
flutter transport (always small here unless driven by reconnection, \ie,
a current gradient).  The energy in the disturbances evolves according to
$$\ptt{}{1\over 8\pi}\avg{\abs{\ptb\Bperp}^2}
	- \div\LP {1\over 4\pi}\ptt{\Afl}\dpp\Afl\RP
	= {1\over n_e e}\avg{\Jfl\dpl\pefl}
	- \avg{\Jfl\dpl\phifl}
	- \avg{\Jfl\ptb R_{ei}}
	\eqno\eqname\eqbflenergy$$
keeping $\ptb R_{ei}$ general as it also contains parallel
gradients of disturbances (cf.~Eq.~\eqrei).


\midinsert
\vskip -10 pt
$$\psboxto(15 true cm;0cm){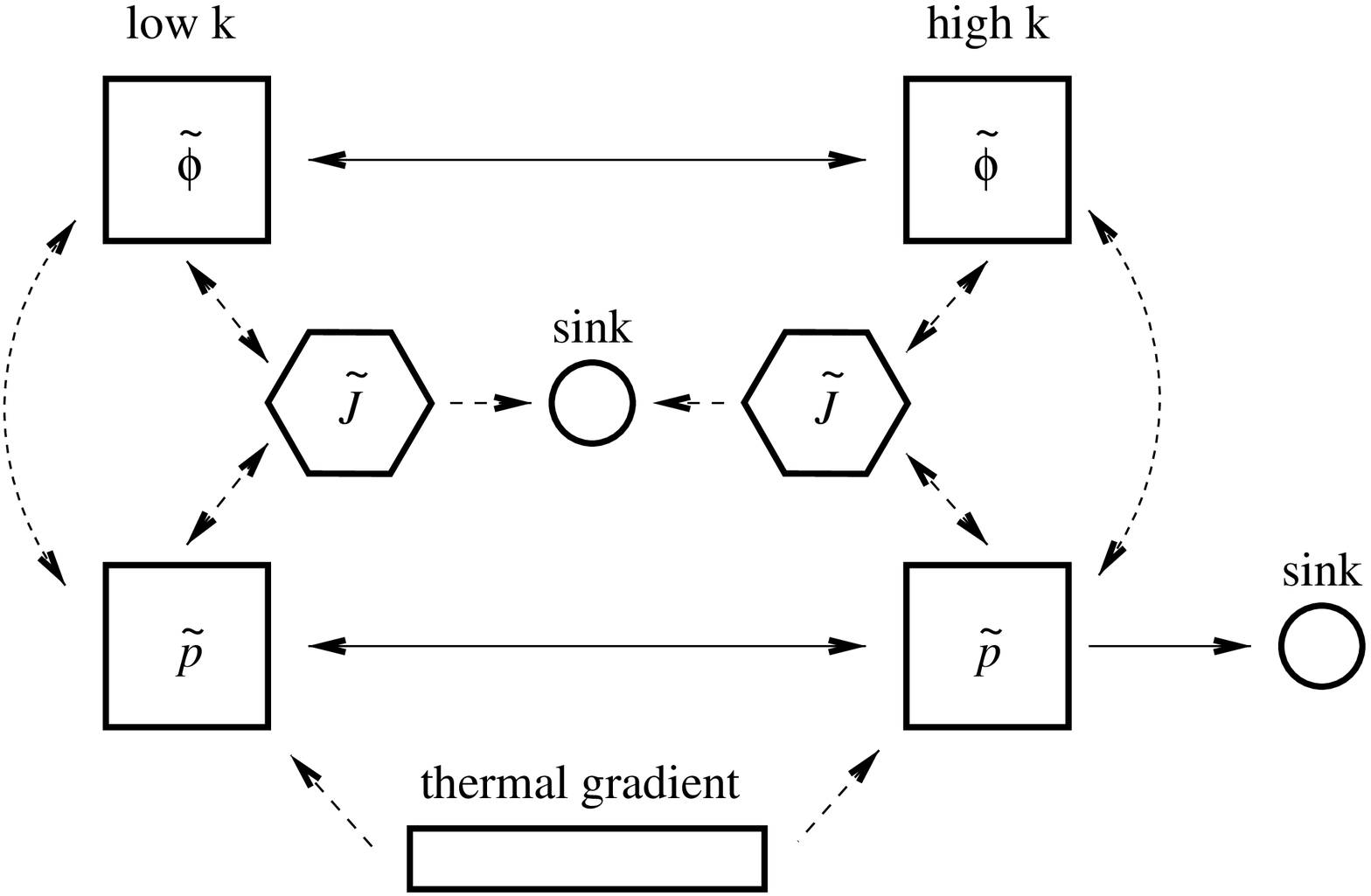}$$
{\hfill\vbox{\hsize=12cm \baselineskip 13 pt \noindent
{\secfnt Figure \figturbtrans.} 
Energy transfer diagram for the disturbances in drift wave turbulence.
Large and small scales are indentified as low and high $\kpp$,
respectively.  The main state variables are $\phifl$ and $\pefl$ and
$\pifl$, the latter two drawn collectively as $\pfl$.  The transfer
channel $\phifl\fromto\Jfl\fromto\pefl$ maintains drift wave mode
structure, while toroidal compression $\pfl\fromto\phifl$ represents
interchange forcing, especially important for ITG turbulence since the
channel through $\Jfl$ does not directly affect the ions.  The weaker
transfer channel $\pefl\fromto\ufl\fromto\pifl$ representing the sound
waves is not shown.  Transfer between scales proceeds through the
polarisation nonlinearity for $\phifl$ and the ExB advection
nonlinearity for $\pfl$.  The principal sink is diffusive mixing of
$\pfl$ out of the energy producing spectral range by ExB eddies, and the
resistive sink is also shown.  The source is the background pressure
gradient.
}\hfill}

\vskip 20 pt
\endinsert


The key question is what to do with the equations for the disturbed
energies in a transport model.  A reasonable treatment is to note that
over an ensemble average the partial time derivatives are zero.  The
energy content of the disturbances is to be neglected in any case,
and since the wave energy transport terms are all small by $O(\delta^2)$
we drop them as well.
These equations then merely become consistency relations among all the
transfer effects.  Eqs.~(\equflenergy,\eqphiflenergy,\eqbflenergy)
respectively become
$$\avg{\ufl\dpl\pifl} + \mupl\avg{\abs{\dpl\ufl}^2}
	= - n_i M_i\avg{\ufl\vefl}\cdot\grad\upl
		- \avg{\ufl\dpl\pefl}
	\eqno\eqname\equflenergy$$
$$\avg{\pifl\div\vefl} = - n_i M_i\avg{\vefl\vefl}\dotdot\grad\uperp
		+ \avg{\Jfl\dpl\phifl}
		- \avg{\pefl\div\vefl}
	\eqno\eqname\eqphiflenergy$$
$$\avg{\Jfl\ptb R_{ei}} - {1\over n_e e}\avg{\Jfl\dpl\pefl}
	= - \avg{\Jfl\dpl\phifl}
	\eqno\eqname\eqbflenergy$$
which will be used to evaluate their the left sides in the equations for
$p_i$, $p_i$, and $p_e$, respectively.
In contrast to the
turbulence, whose nonlinear character is sensitive to the average size
of the transfer effects (\ie, the standard deviation)
\cite{\gyro}, 
the profiles sense only the average transfer and transport (\ie, the
mean).  Some of the transfer effects cancel in this fashion, most
notably the adiabatic transfer mechanism $\avg{\Jfl\dpl\pefl}$.  
The transfer dynamics in ExB turbulence is depicted in Fig.~1, with the
nonlinear drift wave and interchange mechanisms highlighted.  All of the
arrows are bi-directional, due to the fact that the mixing and
scattering tendencies in the nonlinearities, which transfer between
scales of motion, impose their quasi-random character on
the linear mechanisms which transfer between
the state variables $\pefl$, $\pifl$, and $\phifl$, in some cases with
the flux variables $\Jfl$ and $\ufl$ as intermediaries.

The transport model
consists of equations for the usual three thermodynamic state variables,
plus two more for the flow quantities, plus one more for the magnetic
flux.  It is written as follows: 
$$\ptt{n_e} + \div\vec\Gamma = 0
	\eqno\eqname\eqnetrans$$
$$\threehalves\ptt{p_e} + \div \vec Q_e
	= \tei + \npl\Jpl^2
	\eqno\eqname\eqpetrans$$
$$\eqalign{\threehalves\ptt{p_i} + \div \vec Q_i
	& = - \tei 
		- \vec R_E\dotdot\grad\uperp
		- \vec R_E{}_\parallel \cdot\grad\upl
	+ n_i M_i \nu_d u^\theta u^\theta 
	+ {\Jpol\over n_e e}\cdot\grad p_i
\cr}
	\eqno\eqname\eqpitrans$$
$$\div {n_i M_i c^2\over B^2}
	\ptt{}\LP\dpp\phi+{\dpp p_i\over n_e e}\RP
	= \div\drift\Big( M_i\vec\Gamma\cdot\grad\uperp
		+ \div \vec R_E 
		+ n_i M_i\nu_d u^\theta\grad\theta \Big)
	\eqno\eqname\eqphitrans$$
$$n_i M_i\ptt{\upl} + M_i\vec\Gamma\cdot\grad\upl
	+ \div \vec R_E{}_\parallel
	+ n_i M_i \nu_d u^\theta b^\theta
	= 0
	\eqno\eqname\equitrans$$
$${1\over c}\ptt{\Apl} = E_L - \npl\Jpl
	\eqno\eqname\eqapl$$
The parallel current is given by Ampere's law,
$$\Jpl = -{c\over 4\pi}\ddpp\Apl
	\eqno\eqname\eqampere$$
The following transport quantities are defined as ensemble averages over
the turbulence:
$$\vec\Gamma = \avg{\nefl\vefl} \qquad\qquad
	\vec Q_e = \threehalves\avg{\pefl\vefl} \qquad\qquad
	\vec Q_i = \threehalves\avg{\pifl\vefl}
	\eqno\eqname\eqfluxes$$
$$\vec R_E = n_i M_i \avg{\vefl\vefl} \qquad\qquad
	\vec R_E{}_\parallel = n_i M_i \avg{\ufl\vefl}
	\eqno\eqname\eqreynolds$$
$$\tei = \avg{\ufl\dpl\pefl} - \avg{\pefl\div\vefl}
		- \avg{\Jfl\dpl\phifl}
	\eqno\eqname\eqwaveheat$$
giving the particle flux, electron and ion heat fluxes, the
perp-perp and perp-parallel Reynolds stresses, and the anomalous heat
transfer, respectively.



\midinsert
\vskip -10 pt
$$\psboxto(10 true cm;0cm){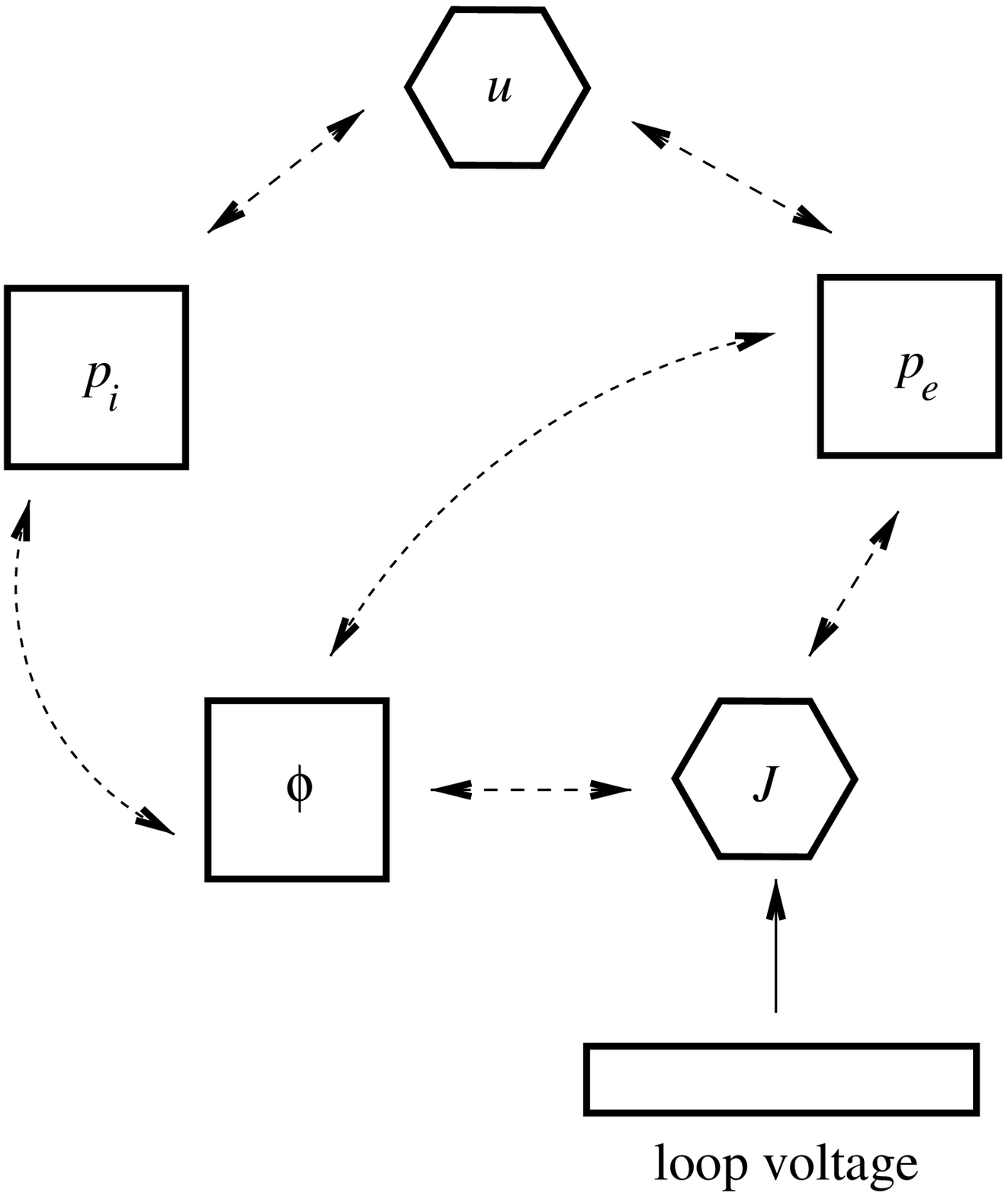}$$
{\hfill\vbox{\hsize=12cm \baselineskip 13 pt \noindent
{\secfnt Figure \figetrans.} 
Energy transfer diagram for the profile quantities in a transport model.
The five quantities representing energy in Eqs.~(\eqpetrans--\equitrans)
are shown, with the transfer channels representing the disturbances.
The three pathways connecting $p_e$ to $p_i$ represent the three terms
in $\tei$ in Eq.~(\eqwaveheat).  The source is the loop voltage.  The
only sink is transport through the outer boundary.
}\hfill}

\vskip 20 pt
\endinsert


\noindent
The perpendicular flow and polarisation current are given by
$$\uperp = \drift\LP\grad\phi+{\grad p_i\over n_e e}\RP
	\eqno\eqn$$
$$\Jpol = \drift\Big( n_iM_i \ptt{\uperp} 
	+ M_i\vec\Gamma\cdot\grad\uperp
	+ \div\vec R_E
		+ n_iM_i\nu_d u^\theta\grad\theta \Big)
	\eqno\eqn$$
and poloidal contravariant components are denoted by the superscript
$\theta$.  As explicit dissipation mechanisms we have the neoclassical
friction $\nu_d$ and the parallel resistivity $\npl$.  The only source
in the model as written is the loop voltage $E_L$.

The most important result of this model is the set of transfer channels
between each of the pressures and the perpendicular and parallel flows,
which are depicted in Fig.~2.
This includes the appearance of anomalous
transfer effects between the electron and ion temperatures, given by 
$\tei$ as defined in Eq.~(\eqwaveheat).  This anomalous transfer is an
essential result of drift wave physics in the turbulence.  The electron
thermal gradient drives ExB motion through the adiabatic and Alfv\'enic
couplings, 
$\pefl\fromto\Jfl\fromto\phifl$.  Dissipation takes place in both
the electron ($\Jfl$) and ion ($\ufl$) channels, which is here
represented in terms of the coupling between the pressures.  More
generally, whichever species has the larger temperature contributes more
to the drive of the turbulence, relaxing the corresponding gradient the
more efficiently.  Rather than a monotonic exchange effect as would be
mandated by dissipative kinetic theory, the anomalous transfer is more a
tendency towards equipartition given a statistical system.
All three of the terms which comprise $\tei$ in
Eq.~(\eqwaveheat) are involved, but $\avg{\Jfl\dpl\phifl}$ is
the strongest for drift wave turbulence, $\avg{\pefl\div\vefl}$ is
strongest for interchange turbulence, and $\avg{\ufl\dpl\pefl}$ is
strongest for ITG turbulence (the pure toroidal ITG transfer goes
through $\avg{\pifl\div\vefl}$, \ie, directly from ion flows to the ion
pressure, not between species).
For strictly cold ions in the absence of flows, the balances among the
disturbances (Eqs.~\equflenergy--\eqbflenergy) would indicate that 
$\avg{\Jfl\dpl\phifl}$ would balance $\avg{\pefl\div\vefl}$ and that 
$\avg{\ufl\dpl\pefl}$ would vanish
and hence that $\tei$ would be zero on average.
However, the presence of both flows as reserviors allows a finite
$\tei$ to exist even for $T_i=0$.  Although this probably depends in
detail on the existence of direct dissipation in the ions (\ie,
viscosity), and therefore a finite $T_i$, it is nevertheless important
to the physics that these transfer channels exist and that they can be
significantly stronger that the classical, collisional transfer usually
considered. 
It is important to note that the drift wave anomalous heat transfer,
$\avg{\Jfl\dpl\phifl}$, representing compression of the polarisation
drift in the ions and parallel currents in the electrons, can be
significant even for $T_i=0$.

The other significant transfer effects are those involving the Reynolds
stresses.  The one most considered is $\vec R_E$, acting
on the perpendicular flow.  But in regions where the parallel flow has a
large gradient the perp-parallel component $\vec R_E{}_\parallel$ also
enters.  It is interesting to note that the direct dissipation in this
model due to $\nu_d$ and $\vec R_E$ acting on the diamagnetic
flow $\ustar$ is exactly cancelled by part of the polarisation transfer,
leaving only the time dependent and advective vorticity contributions,
which constitute a real 
compressional effect.  This does not reduce the computational complexity
however.  But since Eq.~(\eqphitrans) says that $\div\Jpol=0$, the
polarisation current $\Jpol$ may itself vanish in an enclosed domain,
allowing the neglect of the work term involving $\Jpol$ 
entirely.  This only holds if the gradients of $\phi$ and $p_i$ are
set to zero at the outer boundary, however; if it is desired to
incorporate a loss model at that boundary, then one must keep
$\Jpol\cdot\grad p_i$ to conserve energy.  The other unusual effects
regarding the polarisation drift velocity are avoided because of the
neglect of parallel transport in a one-dimensional model, but in a
two-dimensional model they would have to be faced (see the
comments preceding Eq.~\eqenergydef).

It might be tempting to model $\vec R_E$ strictly as flow generation,
expecting it to balance the neoclassical friction in determining the
equilibrium state of $\phi$ \cite{\zonalflows}.
But in many cases the only important
effect on $\phi$ is $\vec R_E$ acting alone; there is evidence from
experimental studies of internal transport barriers that in some
regimes, especially coinjection of neutral beams, (the rotation or
V-cross-B terms in the force balance are stronger than the pressure
gradient and therefore) 
the Reynolds
stress balances itself to zero in response to strong flows 
\cite{\synakowski}, and computations of drift wave turbulence in the
presence of a background ExB vorticity show that the longer wavelengths
are actually driven while the shorter ones are suppressed, so that very
strong flow shear would provide a net drive for the turbulence and
therefore self-deplete \cite{\sfdw}.  Such a self-regulator mechanism is
probably best for modelling $\vec R_E$.  Parallel flow shear is much
less well studied, so the question of what to do for $\vec
R_E{}_\parallel$ is more open.

Current transport codes neglect the partial time derivative in
Eq.~(\eqphitrans) and with it the difficulties in treating the
polarisation transfer; the frictional damping then (or $\vec R_E$
itself) takes over the determination of $\phi$ and the energy in
$\uperp$ is neglected.  But in general $\uperp$ can be as large as
$\upl$, so it is better to keep both flow energies.  Actually, most
transport codes neglect the question of plasma rotation entirely, as
well as the evolution of the magnetic flux, which reduces the set of
equations to Eqs.~(\eqnetrans--\eqpitrans) for $n_e$, $p_e$, and $p_i$,
solely with the particle and energy fluxes, which are easy to model
consistently.  Even in that case, we do still have the anomalous
transfer terms due to the disturbances, and these could become
interesting in several regions of the plasma.  

The existence of anomalous heat transfer has been noted before, in the
context of a quasilinear model for the disturbances
\cite{\waltzglf}.  The more general version presented here in the
context of mean 
field theory incorporates models based upon linear
instabilities as a subset, but it also allows incorporation of the
results from general
turbulence studies, including the strong mode structure changes brought
about by the nonlinear dynamics of the ExB vorticity \cite{\gyro}.
The quasilinear model indicates that $\tei$ is only
important where $\Gamma$ is significant; in turbulence however,
especially electromagnetic turbulence, the robust activity in the
dynamics involving $\dpl\Jfl$ ensures a prominent role for all three of
the contributors to $\tei$ in Eq.~(\eqwaveheat).
Ultimately, further investigation and diagnosis of generalised
turbulence computations will be required to validate any strong
conclusions involving these effects.

\section{V. The Characteristics of Transport Caused
by ExB Turbulence}

In this section we confirm the main points of Section II
concerning the special nature of ExB turbulent transport, compared with
the more familiar random scattering via Coulomb collisions.  

The first point is the simplest: the ExB velocity is nearly divergence
free,
$$\div\vexb\approx 0
	\eqno\eqn$$
and in a homogeneous magnetic field it is divergence free.  For some
laboratory configurations, cylindrical plasma geometry at very low beta
bounded by plates at the ends, for example, this can be taken to be
exact.  The result is something well known from neutral fluid dynamics:
in a divergence free velocity field the pressure does not do work on the
fluid elements.  The transfer of energy between the pressure and the
velocity becomes small.  For interchange turbulence the work done by the
pressure on the ExB velocity divergence is responsible for the principal
forcing effect on the velocity.  But for the profiles this is
overshadowed by the other transfer effects.  A divergence free
transporting velocity is something very different from the model in
collisional kinetics, by which the velocity is given by diffusion,
$$n_e \vv_D = - D(n_e,T_e)\grad n_e
	\eqno\eqn$$
for example.  This velocity is only divergence free for specific
dependences of $D$ on the parameters.  For example, the scale of the
velocity divergence for a constant diffusion is the same as the scale of
the density profile.  But $\vexb$ is exactly or nearly divergence free
according to the magnetic geometry, regardless of what form $n_e(r)$ or
$T_e(r)$ take. 

The second point is the Poynting cancellation, Eq.~(\eqpoyntingcancel).
The presence of $p\vexb$ in the transport with the factor of $3/2$ is
true for total energy, but what we want to know is if that is just a
result of formal manipulation.  We recall that an ideal pressure
equation may be written in terms of a transport term [$(5/2)p\vv$]
and a work term [$\vdl p$], or alternatively in terms of a transport
term [$(3/2)p\vv$] and a divergence [$p\div\vv$].  If the divergence is
small, as we showed for $\vexb$, then the second form is more germane.
But it is even more interesting to know how the transport works for each
piece of the energy rather than the total.  The quantity
$(\phi\jperp+p\vexb)$ appears in the transport equation for the drift
energy, Eq.~\eqendrift.  In modelling transport terms we usually write
down the lowest order forms, which still conserve energy properly.  We
know that the diamagnetic current is at the same order as $\vexb$ if
$n_e e\Eperp$ is comparable to $\grad p$, so $\phi\jperp$ is generally the
same size as $p\vexb$.  In fact they are close enough for the remnant to
be negligible.  A particularly useful form of the Poynting cancellation
turns out to be the one with the diamagnetic current,
$$\div\LP\phi\Jstar+p\vexb\RP
	= \div\LB\drift\grad\LP\phi p\RP\RB
	\eqno\eqn$$
This is a magnetic divergence, and it is now down to the next smallest
row of terms in Eq.~(\eqtotalenergy) as its scale is that of the
magnetic field ($R$ in
toroidal geometry), and the scale of interest is $L_p$.  It is
therefore a fundamental property of ExB flow dynamics that its
contribution to the total energy transport is 
$$\vec Q_e+\vec Q_i = \threehalves\avg{\pfl\vefl}
	\eqno\eqn$$
given that there are no radial equilibrium flows.  The physical point of
this is that the energy due to the fluid perpendicular velocities
presents a transfer term of the form $p\div\vexb$, rather than the more
usual form $\vedl p$, to the thermal energy reservoir, due to the
Poynting cancellation, so that the actual transport is due to the
remaining $(3/2)p\vexb$.  This role of the Poynting cancellation is what
makes the ExB velocity somewhat ``special'' compared to all the others.

The third point concerns magnetic flux diffusion.  It is an interesting
measured property of magnetically confined plasma experiments that the
electron thermal transport is anomalous by more than two orders of
magnitude, but the magnetic flux diffusion is neoclassical
\cite{\neoclres}
(classical
diffusion, modified by toroidal drifts of particles whose parallel
motion is in the long mean free path regime).  If we write the mean
field Ohm's law keeping magnetic flutter, however, 
$${1\over c}\ptt{\Apl} = {1\over n_e e}\div\avg{\pefl\bbfl}
	-\div\avg{\phifl\bbfl} + E_L - \npl\Jpl
	\eqno\eqn$$
we instantly see why this should be true for ExB turbulence,
If the magnetic disturbances are small or, as is more often the case,
uncorrelated, then the anomalous flux diffusion is small; indeed it is
zero for purely electrostatic turbulence.
It is well known that $\avg{\qefl\bbfl}$ contributes
negligibly to transport for drift wave turbulence in either slab or
toroidal geometry, due to self consistency effects between $\phifl$ and
$\tefl$ 
\cite{\dalfloc,\waltz}.
Only when the current gradient is available as an energy
source, as for tearing instabilities,
\cite{\tearing},
do these processes lead to appreciable transport.
It then follows that the anomalous flux diffusion
should be small.  With $\Jpl$ determined by $\Apl$, this holds as well
for the current diffusion.  Anomalous current diffusion can only result
from the ExB nonlinearity in the electron inertia, and this is small
compared to the mechanisms by which currents in a specifically current
carrying plasma are generated.  This is another nice contrast to
classical diffusion, for in that case the flux diffusivity is much {\it
greater} than the particle diffusivity.  Neglecting $\bbfl$, the mean
field Ohm's law is purely diffusive, as we can see from
$$\ptt{\Apl} = E_L + {\npl c^2\over 4\pi}\ddpp\Apl
	\eqno\eqn$$
having inserted Eq.~(\eqampere) to make the flux diffusion explicit.
Comparing the classical diffusivities (assuming a
classical $\npl$), we find 
$${\npl c^2\over 4\pi} = {0.51\over\beta_e}\,{\rho_e^2\nu_e}
	= {0.51\over\beta_e}\,D_e
	\eqno\eqn$$
With $\beta_e\sim 10^{-4}$ in the edge regions of fusion plasmas, this
classical diffusivity is as large as the anomalous electron thermal
diffusivity if not larger.  An interesting consequence is that the MHD
equilibrium in this regime is resistive.  If the pressure gradient
changes during a transport event it is not the case that the
magnetic structure evolves isentropically through its successive
equilibria.  This point has been largely ignored in edge transport
modelling, which takes the magnetic structure to be fixed.

The fourth point is the various mechanisms of ExB flow generation.  Self
consistent flow generation by turbulence goes through the Reynolds stress
\cite{\zonalflows}, 
a fluid dynamical mechanism.  But for closed, toroidal geometry a
sheared ExB velocity profile also means a radial electric field with
radial dependence.  In other words, a finite divergence and hence a
finite charge density.  But within low frequency fluid drift motion at
scales larger than the Debye length, the charge density is always small,
specifically, much less than $n_e e$, the charge density of the
electrons.  Not only that, it is small in all phases of the dynamics,
which we can quickly show.  Writing simple continuity equations for both
fluids using the velocities in Eqs.~(\eqivelocity,\eqevelocity), we find
a charge continuity equation given by
$$\ptt{\rch}+\div\LP\rch\vexb\RP 
	+ \div\LP\Jpl\bunit+\Jstar+\Jpol\RP
	= 0
	\eqno\eqn$$
where $\rch=n_e e - n_i Z e$ is the charge density.  Taking into account
that $\vedl\sim\ppt{}$, the first two terms are the same size.
But the polarisation current can be written next to them,
$$-\ptt{\rch}-\div\LP\rch\vexb\RP -\div\Jpol
	= \div\LP\Jpl\bunit+\Jstar\RP
	\eqno\eqname\eqcharges$$
The polarisation divergence and the charge
density term have the same form, and the polarisation divergence is
larger by a factor of $c^2/v_A^2$, since $\ddpp\phi=4\pi\rch$.  All of
the terms involving $\rch$ are then negligible, and we are left with
$\div\jj=0$ as before.  This underscores the strictness of the
quasineutral character of the dynamics.  It holds pointwise, not just in
an averaged sense.  We then conclude that any mechanism which should
build up an electric field does so under the ExB fluid dynamics and not
due to transport of charges by currents.  The total charge transport
itself must vanish.  This leads to what happens as a result of any
charged particle source or sink mechanisms.  It is quite possible to
have source terms on the right side of Eq.~(\eqcharges), but instead of
accumulating charge we then have a balance between sources and
transport,
$$-\div\Jpol = \div\LP\Jpl\bunit+\Jstar\RP + S_e - S_i
	\eqno\eqn$$
so that the charge content remains small.  This is the high throughput
regime for electric charge, similar to the situation with free energy in
drift wave turbulence where the current transfers thermal and drift free
energy back and forth but the magnetic energy always remains small.  In
the high throughput regime for charges, quasineutrality is maintained
and these source mechanisms act as a torque on the ExB vorticity
\cite{\lackner}.
Taking
charges out of a flux tube causes the flux tube to rotate, since while
the small but finite $\ddpp\phi$ leads to no appreciable charge density,
it the ExB vorticity it does lead to has a significant role.
Ultimately, the source of this angular momentum, as in any case for
perpendicular momentum in fluid drift motion, is the
background magnetic field.

The last point concerns anomalous momentum transport.  With the ExB
vorticity proportional to the small but nonzero charge density,
transport of angular momentum takes place through the Reynolds stress,
which is nothing more than the nonlinear polarisation current.  But in
modelling the transport of this current one should not use an anomalous
conductivity in a simple radial Ohm's law.  Conductivity, or better its
inverse, resistivity, is a friction between the two fluids which
conserves total momentum.  Resistivity leads to particle diffusion, as
the magnetic Lorentz force balancing resistive friction,
$${\vxb\over c} = \npp\jj
	\eqno\eqn$$
with perpendicular resistivity $\npp=m_e\nu_e/n_e e^2$,
gives rise to a collisional drift velocity which acts as a diffusion,
once the MHD equilibrium constraint, $\jxb = c\grad p$, is set in,
$$\vv_D = -{D_e\over T_e}\dpp p
	\eqno\eqn$$
(in the neglect of thermal forces).  But in the momentum equation, the
resistive friction cancels, as
$$\ptt{}n_e m_e\vv + \div\LP\cdots\RP = \vec R_{ie}
	-\grad p_e - n_e e\LP\ee+\vcxb\RP
	\eqno\eqname\eqmotione$$
$$\ptt{}n_i M_i\uu + \div\LP\cdots\RP = -\vec R_{ie}
	-\grad p_i + n_i Z e\LP\ee+\ucxb\RP
	\eqno\eqname\eqmotioni$$
add to become approximately
$$\ptt{}\rho\uu + \div\LP\cdots\RP = - \grad p + \jcxb
	\eqno\eqn$$
The only phenomenon left to transport momentum is the Reynolds stress
(cf.~Eq.~\eqphiprof), which is the dominant effect in the momentum flux
terms we did not explicitly write in Eqs.~(\eqmotione,\eqmotioni).
The JxB force remains as a transfer
mechanism between fluid and Poynting momentum, with the background
magnetic field acting as an anchor.

To summarise, transport of the background profile quantities by small
scale ExB turbulence can still be diffusive, provided the scale of
motion is small compared to $L_p$.  But its fundamental properties are
qualitatively different from those of a kinetic diffusion via random
thermal motions, and these should be taken into account when
constructing transport models.  An example model for one-dimensional
transport including flows and magnetic induction has been given in
Section IV.  Attempts to extend conventional two- and three-dimensional
edge transport models, which are able to treat the parallel fluxes as
well as the radial transport since the profiles are no longer flux
functions in their regime, to incorporate these effects are ongoing
\cite{\rozhansky}.

\par\vfill\eject

{
\parindent 0 pt
\frenchspacing
\parskip=10pt plus 1pt minus 1pt
\def\ref##1.##2\par{\par\hangindent 15pt [##1]##2}
\par\section{References}
\ref \hugill.
	J. Hugill, \nf{23} (1983) 331.

\ref \wootton.
	A. J. Wootton, B. A. Carreras, H. Matsumoto, K. McGuire,
	W. A. Peebles, Ch. P. Ritz, P. W. Terry, and S. J. Zweben,
		\pfb{2} (1990) 2879.

\ref \brag.
	S. I. Braginskii, \revpp{1} (1965) 205.

\ref \anommodels.
	One dimensional semi-empirical transport models up to 1990 are
	reviewed by W. A. Houlberg, D. W. Ross, G. Bateman, S. C. Cowley, 
	P. C. Efthimion, W. W. Pfeiffer, G. D. Porter, D. E. Shumaker,
	L. E. Sugiyama, and J. C. Wiley, \pfb{2} (1990) 2913;
	see also [\hugill] for the earliest efforts.  Some more recent
	efforts are G. Becker, \nf{36} (1996) 1751;
	G. Bateman, A. Kritz, J. Kinsey, A. Redd, and J. Weiland,
		\physp{5} (1998) 1793;
	P. Strand, H. Nordman, J. Weiland, and J. Christiansen,
		\nf{38} (1998) 545;
	G. Vlad, M. Marinucci, F. Romanelli, A. Cherubini,
		M. Erba, V. Parail, and A. Taroni, \nf{38} (1998) 557.
	Two dimensional edge transport models start with the B2 code,
	B. Braams, NET Report No. 68, January 1987 (EUR-FU/XII-80/87/68),
  	A Multi-Fluid Code for Simulation of the Edge Plasma in
	Tokamaks;
	R. Schneider, B. Braams, D. Reiter, H. Zehrfeld, J. Neuhauser,
	M. Baelmans, H. Kastelewicz, and R. Wunderlich,
		{\it Contrib. Plasma Phys.}\vol{32} (1992) 450;
	and the UEDGE code,
	T. Rognlien, P. Brown, T. Campbell \etal, 
		{\it Contrib. Plasma Phys.}\vol{34} (1994) 362;
	T. Rognlien, J. Milovich, M. Resnick, and G. Porter,
		{\it J. Nucl. Mater.}\vol{196-198} (1002) 347.

\ref \boozer.
	A. H. Boozer, \pfb{4} (1992) 2845.

\ref \dalfloc.
	B. Scott, \ppcf{39} (1997) 1635.

\ref \gyro.
	B. Scott, \physp{7} (2000) 1845.

\ref \hasmim.
	A. Hasegawa and K. Mima, \prl{39} (1977) 205; \pf{21} (1978) 87.

\ref \wakhas.
	M. Wakatani and A. Hasegawa, \pf{27} (1984) 611.

\ref \driftorder.
	E. A. Frieman and Liu Chen, \pf{25} (1982) 502.

\ref \refthreehalves.
	T. E. Stringer, \ppcf{33} (1991) 1715;
	D. W. Ross, \ppcf{34} (1992) 137.

\ref \lackner.
	K Lackner, private discussions, 1996.

\ref \hinton.
	F. L. Hinton and C. W. Horton, Jr, \pf{14} (1971) 116.

\ref \redmhd.
	H. Strauss, \pf{19} (1976) 134.

\ref \drake.
	J. F. Drake and T. M. Antonsen, Jr., \pf{27} (1984) 898.

\ref \smolyakov.
	A. Smolyakov, {\it Canadian J. Phys.}\vol{76} (1998) 321.

\ref \tsai.
	S.-T. Tsai, F. W. Perkins, and T. H. Stix, \pf{13} (1970) 2108.

\ref \zonalflows.
	P. Diamond and Y. Kim, \pfb{3} (1991) 1626;
	for a recent review see also 
	P. Terry, {\it Rev. Mod. Phys.}\vol{72} (2000) 109.

\ref\synakowski.
	E. J. Synakowski, S. H. Batha, M. A. Beer, M. G. Bell, 
	R. E. Bell, R. V. Budny, C. E. Bush, P. C. Efthimion, 
	T. S. Hahm, G. W. Hammett, B. LeBlanc, F. Levinton, 
	E. Mazzucato, H. Park, A. T. Ramsey, G. Schmidt, G. Rewoldt, 
	S. D. Scott, G. Taylor, and M. C. Zarnstorff,
	\physp{4} (1997) 1736.

\ref\sfdw.
	B Scott, \ppcf{34} (1992) 1977.

\ref\waltzglf.
	R. E. Waltz, G. M. Staebler, W. Dorland, G. W. Hammett, 
	M. Kotschenreuther, and J. A. Konings, \physp{4} (1997) 2482.

\ref \neoclres.
	M. C. Zarnstorff, K. McGuire, M. G. Bell, B. Grek, D. Johnson,
	D. McCune, H. Park, A. Ramsey, and G. Taylor,
	\pfb{2} (1990) 1852.

\ref \waltz.
	R. E. Waltz, \pf{28} (1985) 577.

\ref \tearing.
	H. Furth, J. Killeen, and M. Rosenbluth,
	\pf{6} (1963) 453.

\ref \rozhansky.
	T. Rognlien, G. Porter, and D. Ryutov, 
		{\it J. Nucl. Mater.}\vol{266\&269} (1999) 654;
	D. Morozov, V. Rozhansky, J. Herrera, and T. Soboleva,
		\physp{7} (2000) 1184.

\par

}

\end